\documentclass[lettersize,journal]{IEEEtran}

\makeatletter
\let\c@lofdepth\relax
\let\c@lotdepth\relax
\makeatother

\usepackage{amsmath,amsfonts}
\usepackage{algorithmic}
\usepackage{algorithm}
\usepackage{array}
\usepackage[caption=false,font=normalsize,labelfont=sf,textfont=sf]{subfig}
\usepackage{textcomp}
\usepackage{stfloats}
\usepackage{url}
\usepackage{xcolor}
\usepackage{verbatim}
\usepackage{graphicx}
\usepackage{cite}
\usepackage{booktabs,float,amsmath,amssymb}
\usepackage{adjustbox}
\usepackage{threeparttable}
\usepackage{multirow}

\newcommand{\rea}{\textcolor{black}}
\newcommand{\reb}{\textcolor{black}}
\newcommand{\rec}{\textcolor{black}}

\newcommand{\reaq}{\textcolor{black}}
\newcommand{\reaqq}{\textcolor{black}}

\begin{document}

\title{USAT: A Universal Speaker-Adaptive Text-to-Speech Approach}

\author{Wenbin Wang, Yang Song, Sanjay Jha
        % <-this % stops a space
\thanks{W. Wang, Y. Song and S. Jha are with the School of Computer Science and Engineering,
University of New South Wales, Kensington, NSW 2052, Australia (e-mail: wenbin.wang@unsw.edu.au, yang.song1@unsw.edu.au, sanjay.jha@unsw.edu.au).}% <-this % stops a space
}

% The paper headers
\markboth{Journal of \LaTeX\ Class Files,~Vol.~14, No.~8, August~2021}%
{Shell \MakeLowercase{\textit{et al.}}: A Sample Article Using IEEEtran.cls for IEEE Journals}

\IEEEpubid{0000--0000/00\$00.00~\copyright~2021 IEEE}
% Remember, if you use this you must call \IEEEpubidadjcol in the second
% column for its text to clear the IEEEpubid mark.

\maketitle

\begin{abstract}
Conventional text-to-speech (TTS) research has predominantly focused on enhancing the quality of synthesized speech for speakers \rec{in} the training dataset. The challenge of synthesizing lifelike speech for unseen, out-of-dataset speakers, especially those with limited reference data, remains a significant and unresolved problem. While zero-shot or few-shot speaker-adaptive TTS approaches have been explored, they have many limitations. Zero-shot approaches tend to suffer from insufficient generalization performance to reproduce the voice of speakers with heavy accents. While few-shot methods can reproduce highly varying accents, they bring a significant storage burden and the risk of overfitting and catastrophic forgetting. \rea{In addition, prior approaches only provide either zero-shot or few-shot adaptation, constraining their utility across varied real-world scenarios with different demands.} Besides, most current evaluations of speaker-adaptive TTS are conducted only on datasets of native speakers, inadvertently neglecting a vast portion of non-native speakers with diverse accents. \rea{Our proposed framework unifies both zero-shot and few-shot speaker adaptation strategies, which we term as ``instant'' and ``fine-grained'' adaptation\rec{s}, respectively, based on their merits.} To alleviate the insufficient generalization performance observed in zero-shot speaker adaptation, we designed two innovative discriminators and introduced a memory mechanism for the speech decoder. To prevent catastrophic forgetting and reduce storage implications for few-shot speaker adaptation, we designed two adapters and a unique adaptation procedure. Additionally, we introduce a new TTS dataset that encompasses 42,000 English utterances from 134 non-native speakers, capturing a wide array of non-native English accents. This dataset is intended to enhance holistic evaluations of adaptive TTS capabilities. Through comprehensive experiments on multiple datasets comprising \rec{both} native and non-native speakers, our approach outperforms contemporary methodologies across various subjective and objective metrics.\footnote{Audio samples can be found at: \url{https://mushanshanshan.github.io/USATDemo/}. The ESLTTS data is available at: \url{https://github.com/mushanshanshan/ESLTTS/}}
\end{abstract}


\section{Introduction}
\IEEEPARstart{R}{ecent} text-to-speech (TTS) technology advancements have increased adoption across numerous applications, encompassing voice assistants and navigational guides. Typically, a TTS model is designed to learn speakers' voices in the training dataset during training and synthesize speech using these voices during inference. Such an approach inevitably affects the model's ability to synthesize speech for unseen speakers, especially speakers with limited reference speech samples \cite{DBLP:conf/icml/ArikCCDGKLMNRSS17, DBLP:conf/iclr/0006H0QZZL21, DBLP:conf/icassp/ShenPWSJYCZWRSA18, DBLP:conf/nips/KimKKY20, DBLP:conf/icml/PopovVGSK21}. Thus, a significant research question in the TTS \rec{domain} is how to clone the voice of unseen speakers with a few speech samples \cite{DBLP:journals/corr/abs-2106-15561}, which is often referred to as ``speaker-adaptive TTS'' or ``voice cloning.''

Speaker-adaptive TTS approaches typically diverge into two paradigms: few-shot and zero-shot speaker-adaptive TTS. \textit{Few-shot speaker-adaptive TTS} \cite{DBLP:conf/interspeech/YuLH0WXLTKL0020, DBLP:conf/interspeech/WangTFYWZ20, DBLP:journals/corr/abs-2210-15868} approaches generally first pre-train a universal TTS model with a multi-speaker dataset, which is subsequently adapted using a handful of utterances from an unseen target speaker, thereby yielding a speaker-specific TTS model that can reproduce the voice of the target speaker. \rea{Compared to zero-shot methods, these approaches can provide more fine-grained adaptation and generate synthesized speech with higher speaker similarities.} However, they necessitate more reference speech data for adapting and extra storage for individualized speaker models \cite{DBLP:conf/iclr/Chen0LLQZL21}. \rec{Moreover}, the adaptation process exposes the system to overfitting and catastrophic forgetting \cite{DBLP:conf/icassp/MossAPGB20}. In contrast, \textit{zero-shot speaker-adaptive TTS} \cite{DBLP:conf/interspeech/TjandraPZK21, DBLP:conf/icassp/ZhaoZWCX22, DBLP:conf/interspeech/Wu00HZSQL22} methods typically involve jointly training a speaker encoder and a universal TTS model. Here, the speaker encoder captures voice features from the target speaker's utterances and generates a speaker embedding, while the universal TTS model synthesizes speech using the target speaker's voice under the guidance of this embedding. \rea{Such frameworks can provide instant adaptation with seconds of reference speech data, which is more flexible than few-shot methods.} However, challenges emerge when the target speaker's voice substantially deviates from the training, e.g., with distinct accents, leading to unsatisfactory speaker similarit\rec{ies} in the synthesized speech \cite{DBLP:journals/corr/abs-2301-02111, DBLP:journals/corr/abs-2306-03509}. \rea{In addition, compared with few-shot methods, zero-shot methods cannot effectively improve synthesized speech quality by utilizing more reference speech. For instance, when the duration of the reference voice data exceeds 15 seconds, zero-shot methods offer a marginal improvement in speaker similarity for the synthesized speech. Finally, most existing speaker-adaptive TTS approaches can only perform either zero-shot or few-shot adaptation. Given the distinct merits and drawbacks of each, a one-size-fits-all adaptation approach would be insufficient for diverse scenarios, especially considering the variations in reference speech duration, quality of the variations for synthesized speech, and speaker accents.} 

\IEEEpubidadjcol

To address the issues of current speaker-adaptive TTS methods, in this work, we propose Universal Speaker-Adaptive Text-to-Speech (USAT). Specifically, our USAT framework is designed \reb{to be capable of }both zero-shot and few-shot speaker adaptation. \rea{Given their unique strengths, we term them \rec{as} \textit{instant adaptation} and \textit{fine-grained adaptation} within USAT. The instant adaptation aims to provide real-time voice cloning, especially when the available reference speech is \reb{only seconds long, or the} target speaker does not have heavy accents. On the other hand, the fine-grained adaptation aims to provide better speaker similarity of the synthesized speech when the available reference speech is longer (e.g., 30 seconds) or the speaker exhibits pronounced accents or non-standard pronunciations.} 

To \rec{tackle} the unsatisfactory speaker similarity issue of instant adaptation and bolster the generalization capacity of USAT's instant adaptation to reproduce as many voices as possible, we incorporate disentangled learning and propose two discriminators for speaker information extraction and timbre conversion process to prevent information leakage. Additionally, we introduce a memory mechanism for the variational autoencoder (VAE), \rec{which can simplify} the distribution of latent speech features. Furthermore, we leverage the representation learning capability of the VAE to refine the speaker encoder. To mitigate potential overfitting and catastrophic forgetting in fine-grained adaptation and to \reb{optimize the storage burden}, we introduce a lightweight flow adapter and phoneme adapter, enabling USAT to synthesize speech in the voices of unseen target speakers with a small number of adapting parameters. 

Furthermore, almost all evaluations for English speaker-adaptive TTS have been restricted to datasets comprising native speakers, primarily due to the absence of high-quality non-native speaker datasets. On the other hand, the global count of non-native English speakers surpasses native speakers by a factor of three and many of them have unique accents that native speakers do not have \cite{nunan2001cambridge}. Therefore, to tackle the issue of the absence of high-quality non-native English speaker datasets, we propose an ESL (English as a Second Language) TTS dataset, comprising 41,000 utterances from 134 non-native English speakers, for more holistic speaker-adaptive TTS evaluation and research in heavily accented scenarios.

Our preliminary version of this work was initially published in \cite{DBLP:conf/interspeech/self}. We have extended our original work as follows. In terms of the model structure, (1) we incorporate a memory-augmented VAE to streamline the posterior distribution, reduce the learning complexity for the downstream module and enhance speech synthesis fidelity; (2) we propose a complementary fine-grained adaptation mode, facilitating the cloning of heavily accented speaker voices; (3) we propose a lightweight flow adapter, which can avoid catastrophic forgetting and overfitting in fine-grained adaptation and significantly reduce speaker-specific model storage space. \rec{Regarding} framework validation, (1) we propose a new dataset specifically designed for speaker-adaptive TTS evaluation; (2) we also introduce additional objective evaluation metrics in instant adaptation evaluation; (3) we conducted a comprehensive experiment to discern the impact of factors such as the adapter structure, position of the adapter, and duration of data used for fine-grained adaptation on the speaker similarity of speech synthesis.

The primary contributions of this paper are:

\begin{itemize}
\item A holistic speaker-adaptive TTS framework, named USAT, can perform both few-shot and zero-shot speaker adaptation, encompassing instant and fine-grained adaptation strategies for diverse real-world speakers’ accents.
\item \rec{With the introduction of the memory-augmented VAE and disentangled representation learning, we have substantially improved the generalization capability of USAT for zero-shot speaker adaptive TTS. Extensive experiments demonstrate that it can outperform all compared approaches in terms of both naturalness and speaker similarity.}
\item \rec{Through our carefully designed few-shot speaker adaptation method, complemented by two plug-and-play flow and phoneme adapters, few-shot speaker adaptive TTS with USAT can achieve competitive speech quality while only adapting 0.5\% to 1.6\% of the parameters compared to other approaches.}
\item We introduce a new ESLTTS dataset for speaker-adaptive TTS evaluations, comprising 41,000 English speech samples from 134 non-native English speakers spanning 31 mother languages, facilitating the evaluation of diverse English accents in real-world scenarios.
\end{itemize}

\section{RELATED WORK}
\label{sec:rel}

\subsection{Zero-shot Speaker Adaptative TTS}

Zero-shot speaker adaptation typically revolves around the joint training of a speaker encoder and a universal speech synthesis model. Here, the speaker encoder extracts a speaker embedding, which represents the unique rhythm and timbre characteristic\rec{s} of the speaker, from untranscribed speech, and then the universal speech synthesis synthesizes speech under the guidance of this embedding \cite{DBLP:conf/icassp/ChienLHHL21, DBLP:journals/corr/abs-2306-07691, DBLP:conf/interspeech/MakishimaSAM22}. Cooper et al. \cite{DBLP:conf/icassp/CooperLYFWCY20} first investigated the efficacy of varied neural speaker embeddings for zero-shot adaptation. They deduced that learnable dictionary encoding-based \cite{DBLP:conf/odyssey/CaiCL18} speaker embeddings \rec{enhanced} speaker similarity and naturalness in synthetic speech, albeit with potential overfitting risks. Xin et al. \cite{DBLP:conf/interspeech/XinSTKS20} expanded speaker adaptation across languages, leveraging a domain adversarial neural network. This network encoded voice attributes into a language-agnostic space, \rec{thus} enhancing naturalness and speaker similarity during target language synthesis. YourTTS \cite{DBLP:conf/icml/CasanovaWSJGP22} also adopted a cross-language adaptation strategy. They introduced an additional language embedding as input to handle different languages and a speaker consistency loss to refine speaker similarity. Kim et al. \cite{DBLP:conf/interspeech/KimJCALK22} explored the correlation between \rec{the} training dataset size and the synthesized speech quality within zero-shot adaptation, noting enhanced naturalness and speaker similarity with expansive pre-training datasets. VALL-E \cite{DBLP:journals/corr/abs-2301-02111} presented a novel speaker-adaptive TTS framework. They leverage\rec{d} a large language model as the acoustic model and utilize\rec{d} the audio codec representation as the learning target. The experimental results show that VALL-E significantly improved the naturalness and speaker similarity of synthesized speech. At the same time, they also found that although the total length of the training data set was expanded to 60,000 hours, it was still challenging to clone the voice of a speaker with a heavy accent. NaturalSpeech 2 \cite{DBLP:journals/corr/abs-2304-09116} employs latent diffusion models, which show powerful generative ability in the computer field \cite{DBLP:journals/tgrs/ZhangLLGSJ24}, to achieve strong in-context learning capabilities for speech synthesis. It utilizes continuous vectors instead of discrete tokens, employs diffusion models for non-autoregressive learning, and incorporates speech prompting mechanisms to facilitate in-context learning. The experimental results show that the speech synthesized by NaturalSpeech 2 \rec{achieves} similar naturalness and speaker similarity to ground-truth speech. \rea{However, previous zero-shot based methods ignore the data distribution shift between unseen and seen speakers, leading to a significant performance gap in the model between seen and unseen speakers. Our method leverages feature disentanglement learning to improve the model's generalization performance and reduce this gap.}

\subsection{Few-shot Speaker Adaptative TTS}

Most few-shot speaker-adaptive TTS approaches \rec{begin} by training a universal TTS model on a multi-speaker dataset and subsequently adapting this model with a handful of samples from an unseen target speaker. This adaptation generates a speaker-specific TTS model, which is capable of synthesizing speech in the target speaker's voice \cite{DBLP:conf/icassp/InoueHAHY020}. Traditionally, HMM-based TTS approaches employ maximum likelihood linear regression for such few-shot adaptations \cite{DBLP:conf/ssw/TamuraMTK98, DBLP:conf/icassp/YamagishiTMK04}. With deep learning-based TTS, Arik et al. \cite{DBLP:conf/nips/ArikCPPZ18} proposed the first few-shot speaker adaptation approach. They pre-trained the Deep Voice 3 \cite{DBLP:conf/iclr/PingPGAKNRM18} in the LibriTTS dataset \cite{DBLP:conf/interspeech/ZenDCZWJCW19} and then adapted it \rec{to} the VCTK dataset \cite{DBLP:conf/ococosda/VeauxYK13}. In a subsequent study, Yang et al. \cite{DBLP:conf/interspeech/YangH20} investigated the influence of the pre-training dataset on the eventual voice quality, emphasizing the significance of using balanced speech datasets from diverse geographical locales. AdaSpeech \cite{DBLP:conf/iclr/Chen0LLQZL21} tackled the surge in computational and storage demands by proposing acoustic condition modeling combined with an innovative conditional layer normalization. This ensured that only \rec{a minimal number of} parameters required fine-tuning, preserving the adaptation quality.  Meta-TTS \cite{DBLP:journals/taslp/HuangLLCL22} employed meta-learning, specifically model-agnostic meta-learning, for speaker adaptation, encompassing dual optimization loops for efficient adaptation.  Finally, AdaSpeech2 \cite{DBLP:conf/icassp/Yan0LQZSL21} and UnitSpeech \cite{DBLP:journals/corr/abs-2306-16083} introduce\rec{d} a mel-spectrum encoder and a unit encoder for extracting semantic information in the speech to replace the text data required in the adaptation process, thus enabling the adaptation process \rec{using} untranscribed speech data by substituting linguistic text features with hidden speech features. \rea{However, previous few-shot approaches fine-tune either the entire or a subset of the pre-trained models, which leads to issues of overfitting and catastrophic forgetting and burden for storing speaker-specific models. In contrast, our approach keeps all pre-trained parameters frozen and only fine-tunes multiple subsequent inserted adapters, which successfully avoids these issues.}

\section{Universal speaker-adaptive Text-to-Speech}
\label{sec:usat}

\subsection{Overview}

\begin{figure}[h]
\centering
\includegraphics[width=0.48\textwidth]{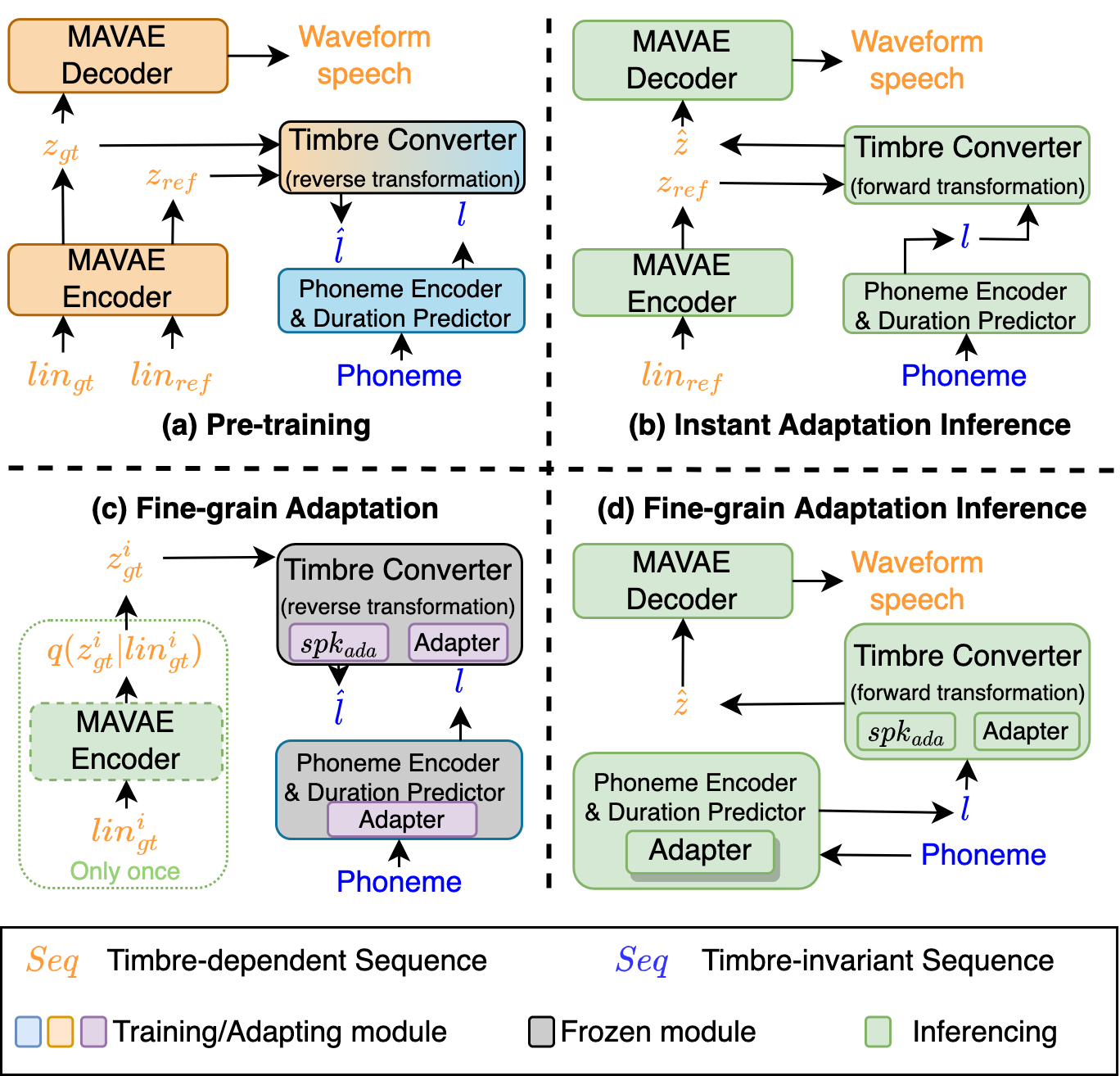}
\caption{The training and inference procedures of USAT's instant and fine-grained adaptation. For clarity in the illustration, the diagram omits some data flows from the timbre converter to the duration predictor.}
\label{ov}
\end{figure}

\rea{As shown in Figure \ref{ov}a, USAT comprises three modules: a \textit{Memory-Augmented Variational Autoencoder (MAVAE)}, a \textit{Timbre Converter}, and a \textit{Phoneme Encoder} paired with a \textit{Duration Predictor}.
The role of the \textit{MAVAE} is two-fold. First, it seeks to derive a frame-level latent speech representation, denoted as \(z\), from the input linear spectrogram. Secondly, it aims to reconstruct the waveform audio from this extracted speech representation. On the other hand, the \textit{Phoneme Encoder} and \textit{Duration Predictor} encode the input phoneme sequence, consequently projecting this onto a frame-level, timbre-invariant phoneme representation, denoted by \(l\). 
The \textit{Timbre Converter} is the core module of the USAT, functioning as a bridge between the timbre-dependent latent speech representation and the timbre-invariant phoneme representation. By providing the target speaker's timbre information via the reference speech representation, denoted as \(z_{ref}\), this converter can perform two inverse transformations: (1) In the pre-training phase, the reverse transformation (expressed as \(z_{gt} \rightarrow \hat{l}\)) disentangles and removes the timbre information from the ground-truth speech representation, producing a timbre-invariant phoneme representation. (2) During the inference stage, the forward transformation (expressed as \(l \rightarrow \hat{z}\)) fuses the timbre information with the phoneme representation, yielding a timbre-dependent speech representation.}

\rea{As illustrated in Figure \ref{ov}a, during the pre-training phase, for each training iteration, linear spectrograms from two utterances of the same speaker are selected as input. One \rec{serves} as the training objective, while the second acts as a reference. 
Then, two speech representations, \(z_{gt}\) and \(z_{ref}\), are derived from these spectrograms. 
The representation \(z_{gt}\) is employed to train the MAVAE decoder for audio reconstruction, and both representations serve as inputs for the timbre converter. The timbre converter aims to decouple and eliminate timbre information in \(z_{gt}\) according to \(z_{ref}\), thereby generating a predicted timbre-invariant phoneme representation, denoted as \(\hat{l}\). Concurrently, the phoneme sequence corresponding to the target utterance is fed into the phoneme encoder to produce a phoneme representation \(l\). The training objective of the timbre converter and phoneme encoder is to minimize the KL divergence between \(\hat{l}\) and \(l\). After this pre-training, the model exhibits sufficient in-context learning capabilities, facilitating instant speaker adaptation with just a few seconds of reference speech. 
In the instant adaptation inference stage, as shown in Fig. \ref{ov}b, the model requires two inputs: a reference utterance and a phoneme sequence. The MAVAE first extracts the reference speech representation from the reference utterance. Simultaneously, the phoneme encoder synthesizes a phoneme representation from the provided phoneme sequence. After that, the timbre converter infuses the phoneme representation with timbre information in the reference speech representation to generate a speech representation denoted as \(\hat{z}\). Finally, the MAVAE decoder reconstructs waveform speech from \(\hat{z}\). It's worth noting that the phoneme encoder and duration predictor inherit the architectural essence from VITS's text encoder and its stochastic duration predictor \cite{DBLP:conf/icml/KimKS21}.}

\rea{When there are sufficient reference utterances of the target speaker, for example, tens of seconds, or when the speaker manifests a distinct accent which the instant adaptation cannot reproduce well, the fine-grained adaptation strategy can be employed to enhance the speaker similarity of synthesized speech. During the fine-grained adaptation stage, as shown in Fig. \ref{ov}c, MAVAE begins by extracting the posterior distributions of speech representations from all available reference utterances, represented as \(q(z^{i}_{gt}|lin^{i}_{gt})\), where \(i\) denotes the i-th reference utterance. After this extraction, MAVAE is no longer involved in the adaptation process. Subsequently, the parameters of both the timbre converter and the phoneme encoder are frozen, and multiple trainable adapters are inserted into them. In parallel, an adaptive speaker embedding, denoted as \(spk_{ada}\), is derived from the reference speech representations, replacing \rec{the} reference speech for providing speaker information to the timbre converter. These newly inserted adapters and the adaptive speaker embedding are subsequently fine-tuned on all reference speech according to the adaptation loss. Note that the speech representation \(z_{gt}\) is sampled from the posterior during each fine-tuning iteration. The fine-grained adaptation inference, as shown in Fig. \ref{ov}d, is similar to the instant adaptation inference process but with two crucial differences: (1) the adaptive speaker embedding replaces the reference speech representation in providing speaker information, and (2) all adapted adapters participate in the inference process. The details of all adapters and the adaptation process are provided in Section \ref{fa}.}

\subsection{Instant Adaptation and Relevant Modules}
\label{ia}

The principal objective of the USAT's instant adaptation is to endow the model with robust generalization capabilities, enabling it to reproduce the voice of the target speaker given just a few seconds of reference utterance in most scenarios. Furthermore, it also aims to set a foundational pre-trained model for USAT's fine-grained adaptation. Fig. \ref{ov}a and Fig. \ref{ov}b visually represent USAT's instant adaptation pre-training and inference phases. 

\subsubsection{Memory-Augmented Variational Autoencoder (MAVAE)}

\begin{figure}[h]
\centering
\includegraphics[width=0.48\textwidth]{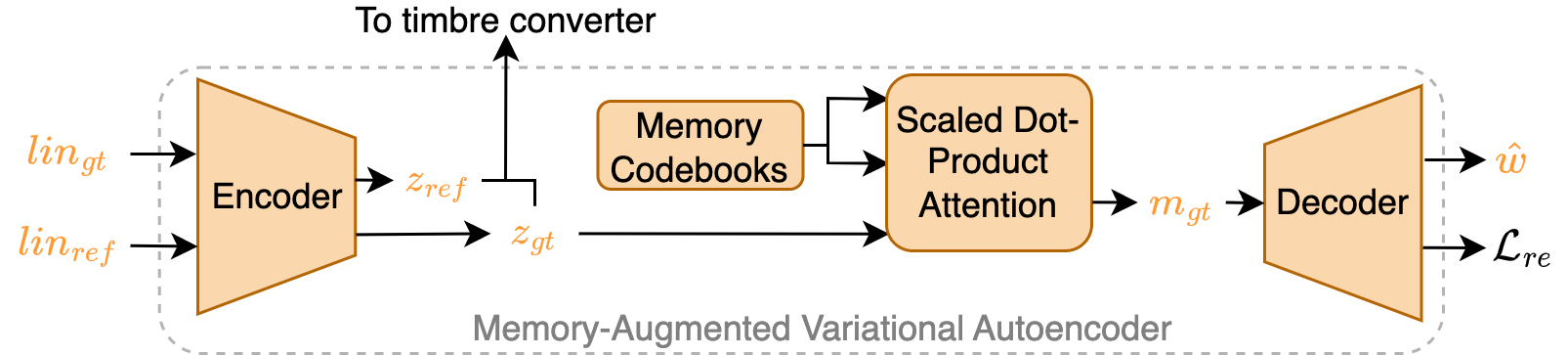}
\caption{The architecture of \rec{the} memory-augmented variational autoencoder.}
\label{mavae}
\end{figure}

\rea{Directly training \rec{a} vanilla VAE with reconstruction loss may lead to a complex posterior distribution \cite{DBLP:journals/corr/abs-2205-04421}.} This complexity challenges downstream modules that leverage this posterior distribution as the learning objective. To streamline the posterior distribution, we introduce MAVAE. As shown in Fig. \ref{mavae}, in MAVAE, \reb{the hidden speech feature \(z_{gt}\), which is sampled from the speech posterior distribution,} serves as a memory codebook query. Subsequently, the output from the memory module, denoted as \(m_{gt}\), is used for the speech waveform reconstruction. This process is mathematically expressed as follows:
\begin{equation}
    z_{gt} \sim q(z|lin_{gt}, \theta_{enc})
\end{equation}
\reaq{
\begin{equation}
    m_{gt} = \left[Softmax(\frac{z_{gt}W_Q(MW_K)^\top}{\sqrt{d}})MW_V \right]W_O
\end{equation}
}
\begin{equation}
    \hat{w} = Dec(m_{gt})
\end{equation}
where \(lin_{gt}\) stands for the input ground-truth linear spectrogram,  \(\theta_{enc}\) denotes the parameters of the MAVAE encoder, \(M\) denotes the learnable memory codebook and the \(W_Q\), \(W_K\), \(W_V\) \reaq{and \(W_O\)} are learnable attention parameters. The term \(d\) signifies the hidden dimension of \(m_{gt}\), \(\hat{w}\) denotes the reconstructed speech. The reconstruction loss of MAVAE \(L_{re}\) is the mean absolute error between the mel-spectrograms of the ground-truth speech and the reconstructed speech:
\begin{equation}
    \mathcal{L}_{re} = \Vert w_{mel} - \hat{w}_{mel} \Vert_1
\end{equation}
where \(w_{mel}\) and \(\hat{w}_{mel}\) represent the mel-spectrogram of the ground-truth and synthesized speech, respectively. 

\subsubsection{Timbre Converter}

\begin{figure}[h]
\centering
\includegraphics[width=0.48\textwidth]{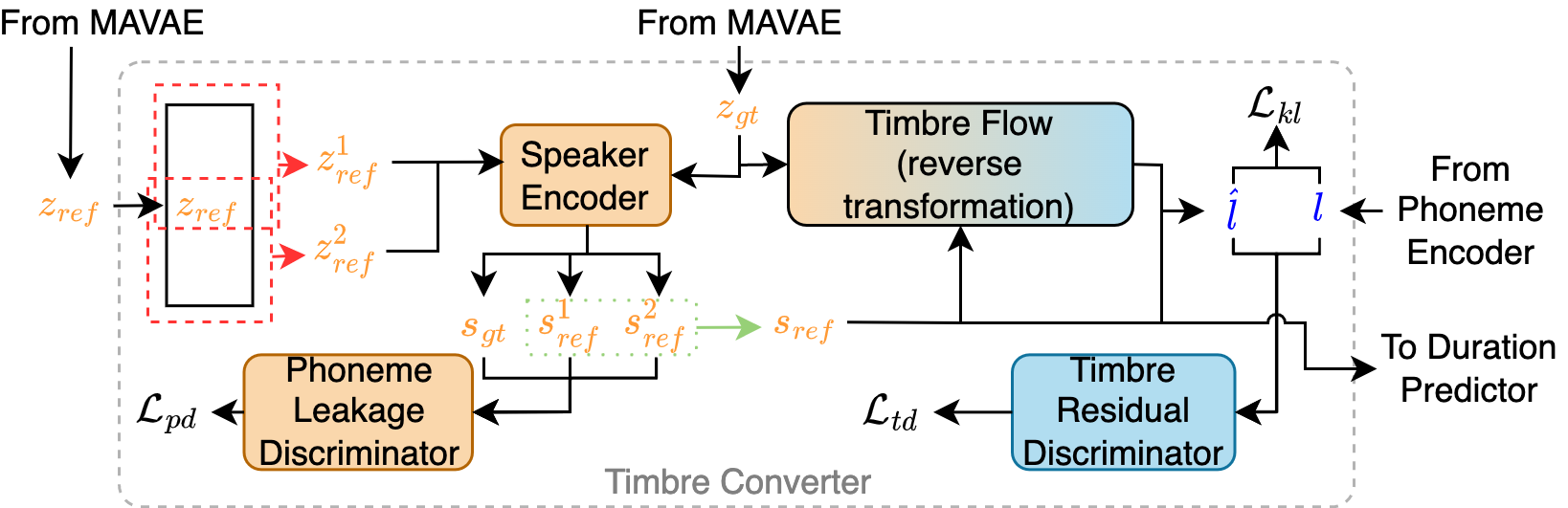}
\caption{The architecture of \rec{the} timbre converter.}
\label{timbre}
\end{figure}

\rea{The timbre converter is the core module for USAT. As shown in Fig. \ref{timbre}, it mainly comprises four modules: \rec{the} speaker encoder, timbre flow and two corresponding discriminators.}

\paragraph{Speaker Encoder}
\label{se}

To enable instant adaptation, we embed a speaker encoder to extract speaker embeddings from the reference utterance. \reaqq{Our method is similar to the global prosody modeling approach introduced in \cite{DBLP:conf/icml/Skerry-RyanBXWS18}, which extracts fixed-dimensional prosody embeddings from the entire reference speech. This is distinct from the local prosody modeling approach in\cite{DBLP:conf/interspeech/KlimkovRRD19}, which begins by obtaining phoneme boundaries through forced alignment and then acquires prosody embeddings for each phoneme by aggregating averages. Our approach offers a simpler training and inference pipeline, without the necessity to perform forced alignment to obtain phoneme boundaries.} \rea{Specifically, the speaker encoder is based on the ECAPA-TDNN \cite{DBLP:conf/interspeech/DesplanquesTD20} since its architecture performs well in capturing \reb{multi-granular} speaker information.} Modifications are made to the original ECAPA-TDNN by substituting its classifier layers with several feedforward layers, enabling the extraction of speaker embeddings. 
\rea{As shown in Fig. \ref{timbre}, during each iteration of the instant adaptation pre-training phase, the reference speech representation will be sliced into two sub-representations with a certain overlap in the time dimension, denoted as \(z_{ref}^1\) and \(z_{ref}^2\). }
The length of this temporal overlap is dictated by the hyperparameter \(\lambda\). Following that, the speaker encoder extracts two speaker embeddings, \(s_{ref}^1\) and \(s_{ref}^2\) from \(z_{ref}^1\)  and \(z_{ref}^2\), respectively. Finally, one of these speaker embeddings is chosen randomly as input for the timbre flow and duration predictor since one embedding already contains enough speaker information for speech synthesizing. Still, both speaker embeddings serve as inputs for the phoneme leakage discriminator.

\paragraph{Phoneme Leakage Discriminator}

Since the hidden representation \(z\) contains highly entangled speaker-relevant (e.g., timbre) and speaker-irrelevant (e.g., linguistic) information, it is challenging for the speaker encoder to disentangle this representation and retain only speaker-relevant information as speaker embedding. To tackle this challenge and prevent contamination with speaker-irrelevant information, particularly linguistic information that could impair the generalizability of the model \cite{DBLP:conf/ijcai/0001LLOQ21}, we propose the \textit{phoneme leakage discriminator}. \rea{This discriminator \rec{is designed to recognize} leaked linguistic information and produce an auxiliary loss,} thereby boosting the speaker encoder’s ability to disentangle speaker-relevant and speaker-irrelevant information. Specifically, during each training iteration, an auxiliary speaker-embedding $s_{gt}$ is extracted from the ground-truth speech by the speaker encoder, combined with $s^1_{ref}$ and $s^2_{ref}$, extracted as described in Section \ref{se}. \reaq{Thus, we can form two contrasting embedding pairs: $[s^{1}_{ref}, s^{2}_{ref}]$ and $[s_{gt}, s^{2}_{ref}]$. The former pair are derived from \(z_{ref}^1\) and \(z_{ref}^2\), which are two segments from the same utterance with overlapping in time dimension. On the other hand, the second pair, originating from \(z_{gt}\) and \(z_{ref}^2\), represents speech features from two utterances by the same speaker. Given that speaker characteristics are relatively time-invariant compared to the variation of linguistic information over time, the primary difference between the two pairs of contrasting embeddings hinges on the potential for overlapping linguistic information. If the speaker encoder is prone to linguistic information leakage, this overlap is likely in the first pair but highly unlikely in the second pair. If a well-trained discriminator fails to determine which pair of speaker embeddings contains more leaked information, it suggests that the leakage can be ignored.} The adversarial penalty \(\mathcal{L}_{se}\) and the training loss for phoneme leakage discriminator \(\mathcal{L}_{pd}\) are defined as follows:
\begin{equation}
\mathcal{L}_{pd} = \mathop{\mathbb{E}}_{s^{1,2}_{ref},s_{gt}}(\mathcal{D}_{p}(s_{gt}\oplus s^{2}_{ref})-1)^2+(\mathcal{D}_{p}(s^{1}_{ref}\oplus s^{2}_{ref}))^2
\end{equation}
\begin{equation}
\mathcal{L}_{se} =\lambda_{se} \mathop{\mathbb{E}}_{s^{1,2}_{ref}}(\mathcal{D}_{p}(s^{1}_{ref}\oplus s^{2}_{ref})-1)^2
\end{equation}
where \(\lambda_{se}\) is a parameter that adjusts the weight of \(\mathcal{L}_{se}\), \reaq{\(\oplus\) denotes the concatenation of vector,} \(\mathcal{D}_{p}\) refers to the phoneme leakage discriminator, a feedforward neural network.

\paragraph{Timbre Flow}
\rea{\rec{To bridge the information gap about timbre between} timbre-dependent sequence and timbre-invariant sequence, we adopt a normalizing flow-based timbre flow to bridge them.} Given the speaker embedding as an auxiliary input, this flow can execute lossless bidirectional transformations, categorized into forward and inverse transformations. The inverse transformation, represented as \(z_{gt} \rightarrow \hat{l}\) in Fig. \ref{timbre}, can eliminate the timbre information in the speaker representation and generate a timbre-invariant phoneme representation. Conversely, the forward transformation, depicted as \(l \rightarrow \hat{z}\) in Fig. \ref{timbre}, can imbue the timbre information \rec{into the} phoneme representation and synthesize timbre-dependent speech representation. The divergence between the generated timbre-invariant phoneme representation and the phoneme representation generated by the phoneme encoder, represented as \(\mathcal{L}_{KL}\), is evaluated using KL divergence, and it can be formulated as follows:  
\reaqq{
\begin{equation}
\mathcal{L}_{kl} = \int p(z_{gt}^{(1:T)}|lin_{gt};\theta) \cdot \frac{p(z_{gt}^{(1:T)}|lin_{gt};\theta)}{q(z_{gt}^{(1:T)}|pho,s_{ref};\theta)} dz
\end{equation}
}
where \(pho\) denotes the input phoneme sequence, \reaqq{\(z_{gt}^{(1:T)}\) denotes the ground-truth speech representation with a length of T frames}, \(lin_{gt}\) denotes the input ground-truth linear spectrogram, \(\theta\) is model parameters.

\paragraph{Timbre Residual Discriminator}

To enhance the speaker similarity of synthetic speech, we propose the timbre residual discriminator for timbre flow. Specifically, since the two transformations of the timbre flow are inverse to each other, enhancing its ability to eliminate the timbre information in the inverse transformation is equivalent to strengthening its ability to imbue the timbre information \rec{into the} phoneme representation in the forward transformation, thereby finally improving the speaker similarity of synthetic speech. To achieve this, we employ the timbre residual discriminator to discern the presence of any timbre information post the inverse transformation, subsequently imposing a corrective penalty on the timbre flow to elevate the quality of its inverse transformation. Given the output sequence from the inverse transformation, denoted as  \(\hat{l}\) in Fig. \ref{timbre}, and the linguistic feature sequence procured from the phonemes, also represented as \(l\) in Fig. \ref{timbre}, the timbre residual discriminator is tasked with identifying sequences that do not contain timbre information. In this context, we integrate the Gradient Reversal Layer (GRL) \cite{DBLP:conf/icml/GaninL15} to invert the gradient's direction. Suppose the timbre converter successfully deceives a proficiently trained timbre residual discriminator. In that case, it stands to reason that its reverse transformation effectively excises most timbre information, while its forward transformation can optimally synchronize timbre specifics with the ground-truth. The timbre flow and timbre residual discriminator will be optimized in different ways according to \(\mathcal{L}_{td}\):
\begin{equation}
\mathcal{L}_{td} = \mathop{\mathbb{E}}_{\hat{l}, l}(\mathcal{D}_{t}(l)-1)^2+(\mathcal{D}_{t}(\hat{l})))^2
\end{equation}
\begin{equation}
\theta\leftarrow\theta-\epsilon(-\lambda_{d}\frac{\partial\mathcal{L}_{td}}{\partial\theta});\ \nu\leftarrow\nu-\epsilon(\frac{\partial\mathcal{L}_{td}}{\partial\nu})
\end{equation}
where \(\theta\) is the parameter of the timbre flow, \(\nu\) is the parameter of the timbre residual discriminator, \(\epsilon\) is the learning rate and \(\lambda_{d}\) is a weight hyperparameter. The timbre residual discriminator mainly consists of multiple Res2Net layers \cite{DBLP:journals/pami/GaoCZZYT21}, an attentive statistics pooling layer \cite{DBLP:conf/interspeech/OkabeKS18} and a classification layer. 

\subsubsection{Phoneme Encoder and Duration Predictor}

\begin{figure}[h]
\centering
\includegraphics[width=0.48\textwidth]{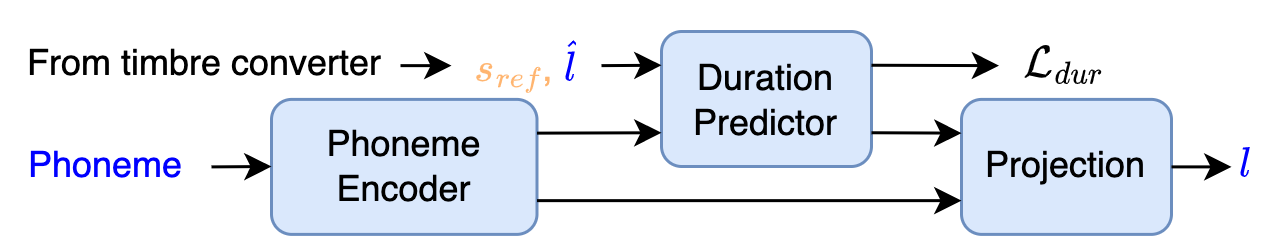}
\caption{The ﬂowchart of \rec{the} phoneme encoder and duration predictor.}
\label{pho}
\end{figure}

\rea{As shown in Fig. \ref{pho}, the phoneme encoder and duration predictor aims to encode the input phoneme sequence into frame-level phoneme representation. The text encoder is a transformer encoder \cite{DBLP:conf/nips/VaswaniSPUJGKP17} with relative positional representation \cite{DBLP:conf/naacl/ShawUV18}. We utilize the stochastic duration predictor \cite{DBLP:conf/icml/KimKS21}, which is based on the monotonic alignment search \cite{DBLP:conf/nips/KimKKY20}, as USAT's duration predictor for end-to-end training. The duration loss \(\mathcal{L}_{dur}\) has the same definition as \rec{that in} the original paper, i.e., the negative variational lower bound of the log-likelihood of the phoneme duration.}

\subsubsection{Final Loss}
The final training loss for the USAT instant adaptation pre-training can be formulated as follows:

\begin{equation}
\mathcal{L}_{train} = \mathcal{L}_{re} + \mathcal{L}_{dur} + \mathcal{L}_{kl} + \mathcal{L}_{se} + \mathcal{L}_{td}
\end{equation}

\subsection{Fine-grained Adaptation}
\label{fa}
The instant adaptation introduces the capability to reproduce most voices. However, it may fail in some scenarios, especially when the speakers have pronounced accents or non-standard pronunciations. Therefore, we further design the \textit{fine-grained adaptation} to address this issue.

\subsubsection{Adaptation and Inference Procedure}

Upon completing the instant adaptation training phase, the pre-trained model has already acquired substantial speaker-adaptive TTS modeling capability. \reb{We refrain from \rec{tuning} the existing learned parameters to take advantage of this \rec{already-learned} capability.} \rea{Specifically, as depicted in Fig. \ref{ov}c, given multiple ground-truth utterances from the target speaker, we first extract the corresponding speech representation posterior distributions from these utterances. Then, we sample multiple speech representations from these distributions and initialize the adaptive speaker embedding by averaging corresponding speaker embeddings:}
\reaq{
\begin{equation}
s_{ada} = \frac{\sum_{i=1}^N(SE(z^i_{gt}))}{N}
\end{equation}
}

\rea{where \(SE\) represents the speaker encoder, this embedding replaces reference speech and speaker encoder to provide the speaker information in the following fine-grained adaptation steps. After that, we remove all discriminators of the timbre flow and freeze all pre-trained parameters in the USAT. \reb{Subsequently, we integrate multiple plug-and-play flow adapters into the timbre flow and the duration predictor and insert multiple phoneme adapters into the phoneme encoder.} Thus, only the parameters of all adapters and adaptive speaker embedding are tunable at the fine-grained adaptation stage.} Finally, these tunable parameters are finetuned on the reference speech utterances according to the adaptation loss. The adaptation loss is defined as:

\begin{equation}
\mathcal{L}_{ada} = \mathcal{L}_{kl} + \mathcal{L}_{dur}
\end{equation}
where the \(\mathcal{L}_{kl}\) and \(\mathcal{L}_{dur}\) are defined the same as those in the instant adaptation. After adaptation, we store the parameters of these adapters and the adaptive speaker embedding as a speaker-specific model, which typically constitutes only 1\% of the number of parameters of the whole USAT model. When inference is required, as depicted in Fig. \ref{ov}d, we integrate the adapters into the pre-trained USAT model and replace the speaker encoder with the adaptive speaker embedding for \rec{inference}.

\subsubsection{Flow Adapter}
\label{flowa}

\begin{figure}[!t]
\centering
\includegraphics[width=0.35\textwidth]{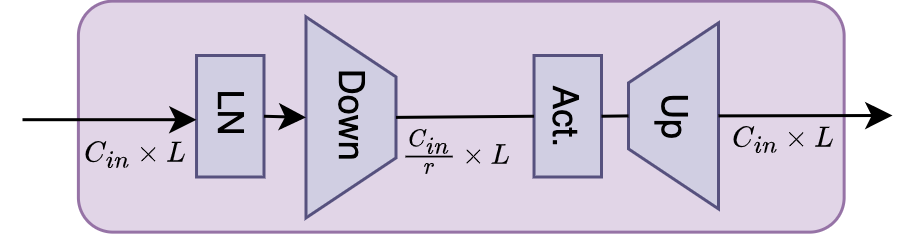}
\caption{The architecture of the flow adapter, \textit{LN} denotes layer normalization, \textit{Act.} represent ReLU activation function, \textit{Down} and \textit{Up} indicate down-projection and up-projection modules. Both projection modules can be either linear or convolution layers depending on the type of adapter.}
\label{fig.adapter}
\end{figure}

\begin{figure}[!t]
\centering
\includegraphics[width=0.45\textwidth]{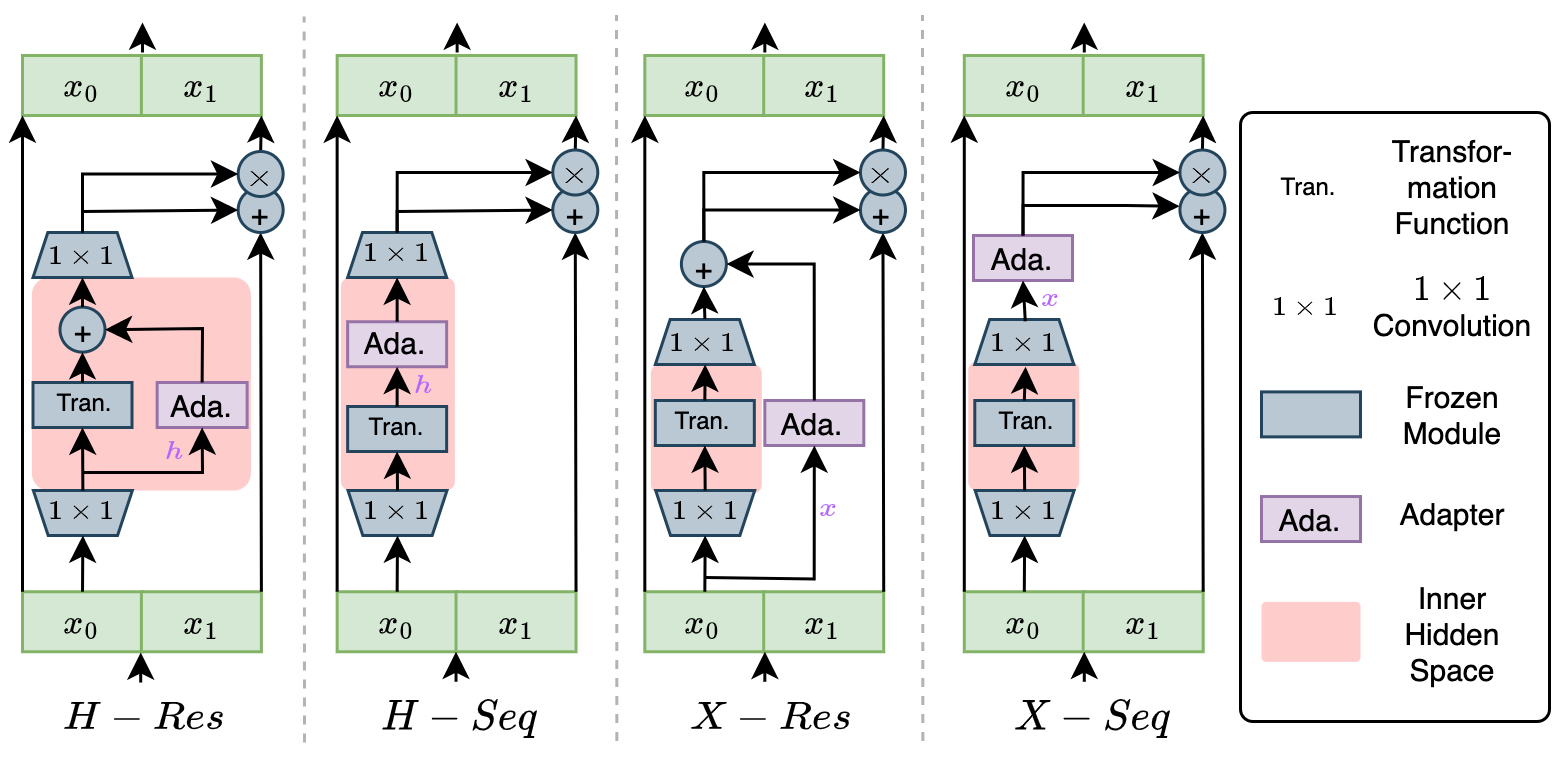}
\caption{The different insertion locations and methods of the flow adapter. \textit{Tran.} denotes the transformation function of the coupling layer, \textit{1x1} denotes the \(1 \times 1\) convolution layer, \textit{Ada.} denotes the flow adapter, \(x_0\) and \(x_1\) denote the input and \(x_0\) and \(x^{\prime}_1\) denote the output. The bottom of the figure indicates the locations of each adapter, \textit{H-*} denotes the adapter is within the coupling layer \textit{X-*} denotes the adapter is alongside the coupling layer, \textit{*-Res} denotes the adapter is residually inserted and \textit{*-Seq} denotes the adapter is sequentially inserted.}
\label{fig.diffadapter}
\end{figure}

The timbre flow and duration predictor are crucial modules influencing the timbre and rhythm of synthetic utterances, and both of them are primarily based on normalizing flow comprised of multiple coupling layers. To this end, we propose the flow adapter for them. As depicted in Fig. \ref{fig.adapter}, the flow adapter first normalizes the input and then down-projects the input \(x \in \mathbb{R}^{c_{in} \times L}\) to \(x' \in \mathbb{R}^{\frac{c_{in}}{r} \times L}\). Following this, a ReLU activation is applied, and \(x' \in \mathbb{R}^{\frac{c_{in}}{r} \times L}\) is up-projected to \(x \in \mathbb{R}^{c_{in} \times L}\). Here, \(r\) represents a hyper-parameter modulating the bottleneck dimension. Through experiments, we discerned that both linear and convolution layers can \rec{effectively} serve as down-projection and up-projection modules for the adapter. Consequently, we evaluated two flow adapter variants: one that employs two convolution layers and another that utilizes two linear layers for the respective projections, named conv-flow adapter and linear-flow adapter, respectively.

Beyond architecture, the position of the flow adapter within the coupling layer also notably impacts adaptation effectiveness. We examined this aspect considering two factors: the adapter's location and insertion mode. As illustrated in Fig. \ref{fig.diffadapter}, considering the typical data flow of the coupling layer, we suggest two insertion locations: one within the transformation function (denoted as \(H\)) and the other adjacent to the transformation function (represented as \(X\)). Additionally, we evaluate two insertion methods: sequential insertion (denoted as \(Seq\)) and residual insertion (indicated as \(Res\)). By integrating these components, we devised four adaptation scheme variants with the coupling adapter: \(H-Res\),  \(H-Seq\),  \(X-Res\) and \(X-Seq\).

\subsubsection{Phoneme Adapter}

\begin{figure}[!t]
\centering
\includegraphics[width=0.35\textwidth]{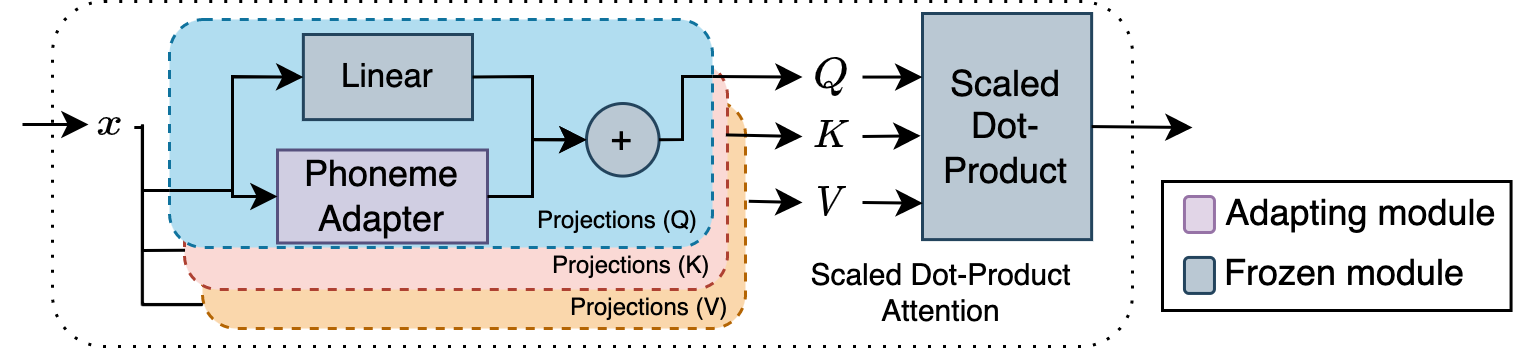}
\caption{\rea{Scaled dot-product attention's architecture after inserting \rec{the} phoneme adapter. The phoneme adapter has the same structure as the linear-flow adapter. All projections (Q, K, V) in the figure have the same structure.}}
\label{pada}
\end{figure}

The pronunciation of English by non-native speakers is often influenced by their native tongues, resulting in deviations from standard pronunciation \cite{nunan2001cambridge}. Considering this is also a speaker's trait, we argue that the speaker-adaptive TTS models should be able to reproduce this. However, the flow adapter in the timbre flow and duration predictor primarily focuses on adapting the speaker's timbre and rhythm, leaving non-standard pronunciation inadequately addressed. To address this, we introduce the phoneme adapter for the phoneme encoder. Given that the phoneme encoder mainly consists of transformer blocks, we follow the common efficient adaptation strategy \cite{DBLP:conf/iclr/HuSWALWWC22} for these blocks, i.e., adapting the scaled dot-product attention weights. \rea{The scaled dot-product attention after \rec{inserting the} phoneme adapter is depicted in Fig. \ref{pada}. The phoneme adapter has the same structure as the linear-flow adapter and is applied to all scaled dot-product attention in the phoneme encoders.}

\section{ESLTTS dataset}
\label{sec:ESLTTS}

\begin{figure}[h]
\centering
\includegraphics[width=0.45\textwidth]{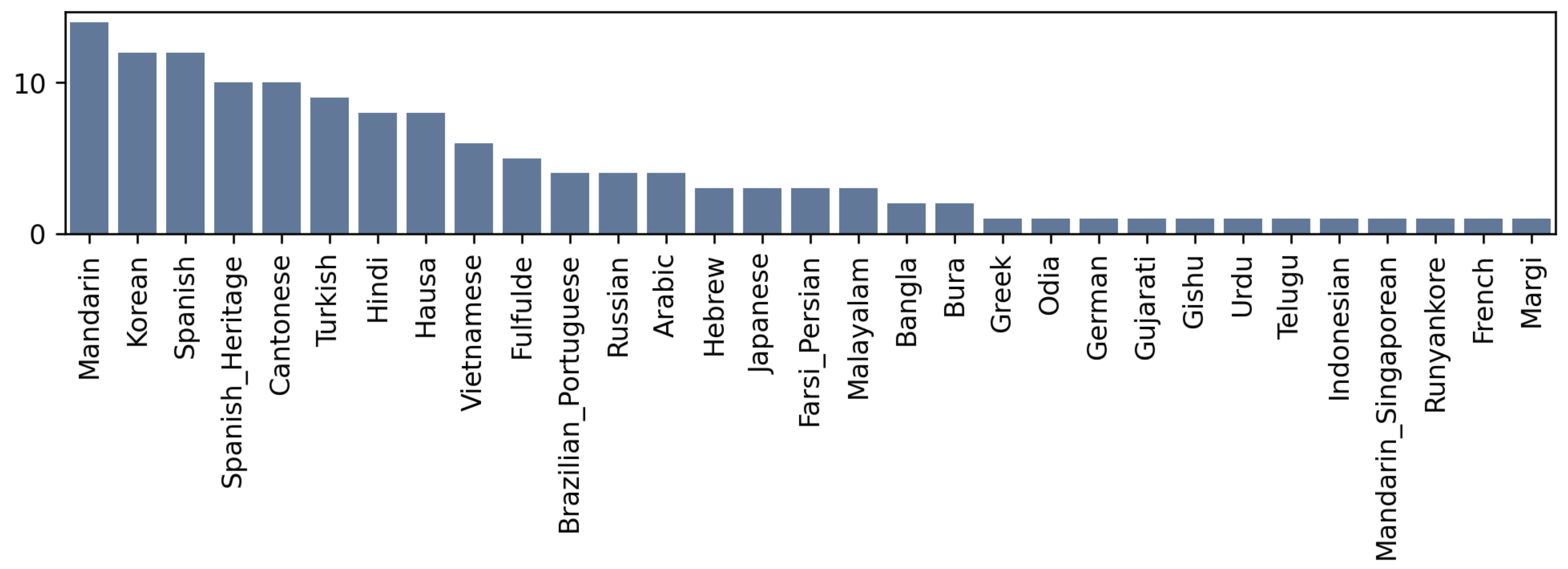}
\caption{\rec{The} number of speakers grouped by each native language in the ESLTTS dataset.}
\label{n_spk_dis}
\end{figure}

With the progress made in speaker-adaptive TTS approaches, advanced approaches have shown a remarkable capacity to reproduce the speaker's voice in the commonly used TTS datasets such as LibriTTS \cite{DBLP:conf/interspeech/ZenDCZWJCW19} and VCTK \cite{yamagishi2019vctk}. However, mimicking voices characterized by substantial accents, such as non-native English speakers, is still challenging. This issue \rec{has been} confirmed in studies by \cite{DBLP:journals/corr/abs-2301-02111, DBLP:journals/corr/abs-2306-03509}. Regrettably, the absence of a dedicated TTS dataset for speakers with substantial accents inhibits the research and evaluation of speaker-adaptive TTS models under such conditions. To \rec{address} this gap, we developed a corpus consisting solely of English utterances from non-native speakers.

\rec{We named this corpus \rec{``English as a Second Language TTS dataset\footnote{Online Available at: \url{https://github.com/mushanshanshan/ESLTTS}}''} (ESLTTS). The ESLTTS dataset consists of roughly 37 hours of 41,000 utterances from 134 non-native English speakers. These speakers represent a diversity of linguistic backgrounds spanning 31 native languages, as depicted in Fig. \ref{n_spk_dis}. For each speaker, the dataset includes an adaptation set lasting about 5 minutes for speaker adaptation, a test set comprising 10 utterances for speaker-adaptive TTS evaluation, and a development set for further research. To ensure the diversity of accents within the ESLTTS dataset, we curated the data from several existing datasets: }

\rec{\textbf{AccentDB \cite{ahamad-anand-bhargava:2020:LREC}} is an English dataset tailored for automatic speech recognition, particularly in scenarios involving non-native accents.}

\rec{\textbf{Google Crowdsourced Speech Corpora \cite{DBLP:journals/corr/abs-2010-06778}} is a corpus collection encompassing multi-language datasets from various continents for developing automatic speech recognition applications. Due to quality and diversity concerns, we selectively used the English subset collected from Africa.}

\rec{\textbf{L2-arctic \cite{zhao2018l2arctic}} is an English speech corpus designed for voice and accent conversion and collected from non-native English speakers.}

\rec{\textbf{Allsstar \cite{bradlow2010allsstar}} is a dataset containing both English and native language utterances from the same speaker, collected from multiple non-native English speakers. It is designed to study the effects of a non-native English speaker’s mother tongue on their English speech, for instance, in terms of speech speed and pronunciation.}

\rec{Given that none of the above source datasets were explicitly designed for TTS, it's unsuitable to employ them for evaluation directly. For example, some utterances have considerable background noise, some have inaccurate ground truth text, and some contain speech from multiple speakers. Therefore, we implemented a series of preprocessing steps to ensure the data's compatibility with TTS applications:}

\begin{itemize}
\item Removed silence from the start and end of each sentence using the \rec{Silero Voice Activity Detector\footnote{Publicly available at: \url{https://github.com/snakers4/silero-vad}}}.
\item Segmented the speech and discarded audio that is less than 0.5 seconds in duration.
\item Excluded speech segments with a signal-to-noise ratio (SNR) of less than 0dB, computed based on the waveform amplitude distribution analysis as detailed in \cite{DBLP:conf/interspeech/KimS08}.
\item Transcribed utterances using the Whisper \cite{DBLP:journals/corr/abs-2212-04356}, calculated the word-level edit distance between the ground truth text and transcription, and removed the utterance whose edit distance is larger than 1.
\item Removed utterances containing modal particles.
\item Removed utterances containing speech from multiple speakers using Pyannote-audio \cite{Bredin2020}.
\item Divided the adaptation set and test set for each speaker.
\item Normalized non-standard terms (e.g., abbreviations) to spoken-style words using a weighted finite-state transducer-based text normalizer \cite{bakhturina22_interspeech}.
\item Downsampled all speech samples to a 24,000 Hz sampling rate for consistency.
\end{itemize}

\section{EXPERIMENTS}
\label{sec:exp}

\subsection{Dataset}

\rec{Our USAT framework is trained on the LibriTTS training set. We subsequently evaluate its instant adaptation capacity on the LibriTTS test set, VCTK, and ESLTTS datasets. Additionally, we conduct fine-grained adaptation evaluation on the ESLTTS dataset.}

\textbf{LibriTTS \cite{DBLP:conf/interspeech/ZenDCZWJCW19}.} LibriTTS is a highly prevalent multi-speaker English TTS dataset, containing 585 hours of speech data sampled at 24,000 Hz from 2456 speakers. There are 354,780 utterances for training and 9,957 utterances for testing. \rec{Following other approaches \cite{DBLP:conf/icml/MinLYH21, DBLP:conf/interspeech/Wu00HZSQL22}, we randomly selected 30 speakers out of the 39 available in the LibriTTS test set and chose 10 random sentences per speaker for evaluation.}

\textbf{VCTK \cite{yamagishi2019vctk}.} The VCTK dataset, another prominent multi-speaker English TTS dataset, comprises 25 hours of speech data sampled at 48,000 Hz, sourced from 109 speakers. The original dataset does not split training and test sets and contains 44,283 utterances. \rec{We also randomly selected 30 out of 109 from the VCTK dataset with 10 randomly selected sentences per speaker for evaluation.}

\textbf{ESLTTS.} As described in Section \ref{sec:ESLTTS}, the ESLTTS dataset is specially formulated for researching and evaluating speaker-adaptive TTS tasks under substantial accent conditions. The dataset consolidates 37 hours of speech data, sampled at 24 kHz, from 134 non-native English speakers. During instant adaptation evaluation, 30 out of 134 speakers are randomly selected from the ESLTTS adaptation set, ensuring the inclusion of at least one speaker from each mother tongue; for each speaker, evaluation is conducted on the speaker's test set. During fine-gained adaptation evaluation, we chose the same 30 speakers, adapted the model using their adaptation set, and evaluated the adapted model using their test set.

\subsection{Compared methods}

\rec{For all the comparison methods, we used their official code and checkpoints for evaluation. Since StyleSpeech, MetaStyleSpeech and YourTTS synthesize speech at 16,000 Hz, we downsampled all synthesized speech to 16,000 Hz during the evaluation.}

\subsubsection{Zero-shot speaker adaptation approaches}
We compare the instant adaptation of USAT with the following zero-shot speaker adaptation approaches:

\rec{\textbf{StyleSpeech\footnote{Available at: \url{https://github.com/KevinMIN95/StyleSpeech}} \cite{DBLP:conf/icml/MinLYH21}} is the previous state-of-the-art zero-shot speaker-adaptive English TTS approach.} It is based on FastSpeech2 \cite{DBLP:conf/iclr/0006H0QZZL21} and introduces the Style-Adaptive Layer Normalization, which aligns the gain and bias of the text input according to the style extracted from the reference speech audio. 

\rec{\textbf{Meta-StyleSpeech\footnote{Available at: \url{https://github.com/KevinMIN95/StyleSpeech}} \cite{DBLP:conf/icml/MinLYH21}} is a variant of StyleSpeech. It introduces meta-learning for StyleSpeech and improves synthetic speech quality by incorporating two discriminators and using episodic training. }

\rec{\textbf{YourTTS\footnote{Available at: \url{https://github.com/Edresson/YourTTS}} \cite{DBLP:conf/icml/CasanovaWSJGP22}} is the currently publicly available state-of-the-art zero-shot speaker-adaptive English TTS approach.} It employs the H/ASP \cite{DBLP:journals/corr/abs-2009-14153} model-based speaker encoder and trains it with a speaker consistency loss to ensure high speaker similarity between synthetic and ground-truth speech. \reaq{We utilized the ``Exp 4'' checkpoint for evaluations.}

\subsubsection{Few-shot speaker adaptation approaches}
Additionally, the fine-grained adaptation of USAT is compared with the following few-shot speaker adaptation approaches:

\textbf{VITS (Full Tuning) \cite{DBLP:conf/icml/KimKS21}} fine-tunes all parameters within a vanilla VITS model during the adaptation stage, which is pre-trained on the LibriTTS dataset. 

\rec{\textbf{UnitSpeech\footnote{Available at: \url{https://github.com/gmltmd789/UnitSpeech}} \cite{DBLP:journals/corr/abs-2306-16083}} is a recently proposed advanced diffusion-based few-shot speaker adaptive TTS approach.} It employs a unit encoder to provide linguistic information to the diffusion-based acoustic module and then fine-tune the decoder for speaker adaptation to the reference speaker without requiring corresponding transcriptions.

\subsection{Evaluation Metrics}

We utilized both subjective and objective evaluation metrics to evaluate synthesized utterances' naturalness and speaker similarity. The details of all metrics are as follows:

\textbf{Naturalness Mean Opinion Score (NMOS) \cite{DBLP:conf/nips/ArikCPPZ18}.} We assemble all synthesized utterances by all approaches along with the ground-truth utterance for each evaluation sample for human evaluation. The participants then assess the naturalness of each utterance on a five-point Likert Scale (1: Completely unnatural speech, 2: Mostly unnatural speech, 3: Equally natural and unnatural speech, 4: Mostly natural speech, 5: Completely natural speech). Each sample is evaluated by 5 participants. We calculate the mean score and confidence interval for each approach. All subjective evaluations were conducted through qualtrics\footnote{\url{https://www.qualtrics.com/au/}} among native English speakers with ethical approval\footnote{UNSW HREAP Executive Approval Number: HC230370}.

\textbf{Speaker Similarity Mean Opinion Score (SMOS) \cite{DBLP:conf/nips/ArikCPPZ18}.}  In parallel with NMOS, the speaker similarity between the synthesized utterance and a random utterance from the same speaker is also evaluated. The assessment is carried out similarly on a five-point Likert Scale (1: Not similar at all, 2: Slightly similar, 3: Moderately similar, 4: Very similar, 5: Extremely similar) by the same participants. We also calculate the mean score and confidence interval for each approach.

\textbf{Word Error Rate (WER) \cite{DBLP:conf/icml/MinLYH21}.} The WER represents the average ratio of errors in synthetic speech transcripts to the ground-truth text and is widely used in the evaluation of automatic speech recognition \cite{WANG2024216}. \rec{A lower WER suggests fewer pronunciation errors in synthesized speech.} We use Whisper \cite{DBLP:journals/corr/abs-2212-04356} for speech transcription.

\textbf{UTokyo-SaruLab Mean Opinion Score (UTMOS) \cite{DBLP:conf/interspeech/SaekiXNKTS22}.} \rec{Following other works \cite{DBLP:journals/taslp/HuangLLCL22}, we also provide predicted NMOS for reference. We utilized the UTMOS for NMOS prediction, which achieved state-of-the-art performance in 10 of the 16 metrics at the VoiceMOS Challenge 2022 \cite{huang22f_interspeech}.}

\textbf{Speaker Embedding Cosine Similarity (SMCS) \cite{DBLP:conf/icml/MinLYH21}.} SMCS quantifies the voice similarity between the synthesized utterance and a random ground-truth utterance from the same speaker by calculating the cosine similarity of speaker embeddings extracted from the respective utterances. \rec{Unlike \cite{DBLP:conf/icml/CasanovaWSJGP22}, which utilized the GE2E \cite{DBLP:conf/icassp/WanWPL18} model to extract speaker embeddings, we utilize a more powerful speaker verification model, TitaNet-L \cite{DBLP:conf/icassp/KoluguriPG22}. It achieves the state-of-the-art equal error rate on the VoxCeleb1 test set in the speaker verification task and hence can better distinguish subtle differences between voices.} 
 
\textbf{Speaker Verification Rate (SVR) \cite{DBLP:conf/icassp/ChienLHHL21}.} In alignment with the goal of speaker-adaptive TTS, i.e., synthesizing speech using the target speaker's voice, \rec{following \cite{DBLP:conf/icassp/ChienLHHL21, DBLP:journals/corr/abs-2308-04177}, we also introduce the SVR metric as an additional reference.} It represents the percentage of synthetic speech successfully verified by the speaker verification system as originating from the same speaker as a random ground-truth speech. \rec{TitaNet-L \cite{DBLP:conf/icassp/KoluguriPG22} is again used as the speaker verification system. It marks two utterances as from the same speaker when SMCS is greater than 0.7.}

\subsection{Implementation Details}

\subsubsection{Pre-training Procedure} We pre-train USAT on the LibriTTS training set. During pre-training, all samples are down-sampled to a sampling rate of 22,050 Hz for the training process. The corresponding phoneme sequences are derived using the Phonemizer \cite{Bernard2021}, a widely used grapheme-to-phoneme toolkit. The input linear spectrogram is transformed using the short-time Fourier Transform with a window length of 1024, a frameshift of 256, and an FFT size of 1024. The values for $\lambda_{se}$ and $\lambda_{d}$ are set to 8, while $\lambda_{ol}$ is set to a range of 20\% to 40\% to prevent the discriminator from overpowering the timbre flow. We train all modules end-to-end for 550k iterations using 8 V100 GPUs with a total batch size of 128. The AdamW optimizer \cite{DBLP:conf/iclr/LoshchilovH19} is employed with $\beta_1=0.8$, $\beta_2=0.99$, and a weight decay of 0.01. The learning rate is initialized to $2\times10^{-4}$, with a decay factor of $\gamma=0.9999$.

\begin{table*}[]
\setlength\tabcolsep{3pt}
\caption{Evaluation Results of Instant Speaker Adaptation in LibriTTS.}
\begin{adjustbox}{width=15cm,center}
\begin{threeparttable}
\label{zsel}
\scriptsize
\begin{tabular}{c|cccccc|cccccc}
\toprule
\(\rm Dataset_{speaker}\)               & \multicolumn{6}{c|}{\(\rm LibriTTS_{unseen}\)}         &\multicolumn{6}{c}{\(\rm LibriTTS_{seen}\)}\\ \midrule
Metric          & NMOS  & SMOS    & SMCS                      & WER(\%) & UTMOS  & SVR(\%) & NMOS    & SMOS     & SMCS                    & WER(\%)    & UTMOS  & SVR(\%)            \\ \midrule
Ground-Truth    & \(4.39\pm0.09\) & -             & 0.894        & 2.3 & 4.09 & 100  & \(4.37\pm0.08\) & -     & 0.890               & 2.4  & 4.13 & 100   \\ \midrule 
StyleSpeech \cite{DBLP:conf/icml/MinLYH21}  & \(3.29\pm0.07\) & \(3.43\pm0.06\)    & 0.672  & 6.1 & 3.34 & 36.2 & \(3.46\pm0.08\)  & \(3.61\pm0.07\) & 0.707   & \boldmath\(3.4\)   & 3.46 & 58.9     \\
Meta-StyleSpeech \cite{DBLP:conf/icml/MinLYH21} & \(3.28\pm0.07\) & \(3.44\pm0.09\) & 0.667  & 6.3 & 3.38 & 36.2 & \(3.47\pm0.07\)  & \(3.59\pm0.08\)  & 0.703  & 3.5  & 3.47 & 55.1      \\
YourTTS \cite{DBLP:conf/icml/CasanovaWSJGP22} & \(3.70\pm0.09\) & \(3.65\pm0.07\)  & 0.702  & 6.0 & 3.69 & 67.8 & \(3.84\pm0.08\)  & \(3.82\pm0.09\)  & 0.741  & 5.8 & 3.67 & 75.1 \\
USAT         & \boldmath\(3.96\pm0.08\) & \boldmath\(3.98\pm0.08\)  & \textbf{0.751}  & \boldmath\(5.9\) & \textbf{3.81} & \textbf{80.1} & \(3.95\pm0.08\) & \boldmath\(4.06\pm0.05\) & \textbf{0.780}   & 5.8 & \textbf{3.80} & \textbf{88.7} \\ \bottomrule
\end{tabular}
\end{threeparttable}
\end{adjustbox}
\end{table*}

\begin{table*}[]

\setlength\tabcolsep{3pt}
\caption{Evaluation Results of Instant Speaker Adaptation in VCTK and ESLTTS.}
\begin{adjustbox}{width=15cm,center}
\label{zsee}
\begin{threeparttable}
\scriptsize
\begin{tabular}{c|cccccc|cccccc}
\toprule
\(\rm Dataset_{speaker}\)               & \multicolumn{6}{c|}{\(\rm VCTK_{unseen}\)}         &\multicolumn{6}{c}{\(\rm ESLTTS_{unseen}\)}\\ \midrule
Metric          & NMOS  & SMOS    & SMCS                      & WER(\%) & UTMOS  & SVR(\%) & NMOS    & SMOS     & SMCS                    & WER(\%)    & UTMOS  & SVR(\%)            \\ \midrule
Ground-Truth    & \(4.42\pm0.07\) & -               & 0.874     & 5.1    & 3.98 & 100  & \(4.26\pm0.09\) & -     & 0.862                & 12.6  & 3.84 & 99.5   \\ \midrule 
StyleSpeech \cite{DBLP:conf/icml/MinLYH21}  & \(3.23\pm0.08\) & \(3.22\pm0.08\) & 0.674 & 13.2   & 3.41 & 34.7  & \(3.43\pm0.09\)  & \(2.52\pm0.08\) & 0.606   & 14.2  & 3.20 & 7.5      \\
Meta-StyleSpeech \cite{DBLP:conf/icml/MinLYH21} & \(3.21\pm0.08\) & \(3.22\pm0.09\) & 0.676 & 13.4   & 3.44 & 31.9 & \(3.42\pm0.09\)  & \(2.65\pm0.08\)  & 0.606  & 15.4  & 3.21 & 7.5       \\
YourTTS \cite{DBLP:conf/icml/CasanovaWSJGP22} & \(3.71\pm0.08\) & \(3.60\pm0.07\) & 0.745 & 11.9   & 3.67 & 81.5 & \(3.78\pm0.06\)  & \(3.01\pm0.09\)  & 0.674  & 14.1 & 3.68 & 32.7 \\
USAT         & \boldmath\(3.84\pm0.07\) & \boldmath\(3.88\pm0.08\) & \textbf{0.753} & \boldmath\(11.2\)   & \boldmath\(3.81\) & \boldmath\(83.8\) & \(3.86\pm0.06\) & \boldmath\(3.22\pm0.09\) & \textbf{0.694}   & \textbf{14.0} & \textbf{3.82} & \textbf{48.4}\\ \bottomrule
\end{tabular}
\end{threeparttable}
\end{adjustbox}
\vspace{-0.2cm}
\end{table*}

\subsubsection{Evaluating the Variables of Fine-grained Adaptation} 
\label{fsada}
We conducted a comprehensive evaluation focusing on various factors that may influence USAT's fine-grained adaptation: (1) For flow adapter insertion method and position, we evaluated all four distinct insertion methods and position combinations of flow adapter described in \ref{flowa}, i.e., \(H-Res\),  \(H-Seq\),  \(X-Res\) and \(X-Seq\). (2) For hyperparameter \(r\) we evaluated two values of the hyperparameter \(r\), i.e., 4 and 8. Also, to maintain consistency in the number of parameters across each adapter, for the adapters \(X-Res\) and \(X-Seq\), we use \(\frac{r}{2}\) instead. (3) For the duration of adaptation sets, we assessed the adaptation effectiveness for two distinct durations: 60 seconds and 300 seconds. The 300-second set corresponds to the original ESLTTS adaptation set, whereas the 60-second set is a subset derived from the original adaptation set. (4) For adaptation steps, we adapted 2,000 steps for each experiment. Thereafter, we analyze the changes in SMCS and SVR every 100 adaptation steps.

\section{EXPERIMENTAL RESULTS}
\label{sec:expres}

\subsection{Instant Speaker Adaptation Evaluation}

\subsubsection{Evaluation in LibriTTS and VCTK} We first evaluate the instant adaptation for USAT in slight to no accent speakers, i.e., speakers in LibriTTS and VCTK. The results are shown in Tables \ref{zsel} and \ref{zsee}, respectively. \reaq{The results demonstrate that USAT surpasses the competing methods in most evaluation metrics. Specifically, for seen speakers in LibriTTS, i.e., the speakers in the LibriTTS training dataset, USAT exhibits improvements of 6.28\% for SMOS, compared to YourTTS.} When evaluating unseen speakers in the LibriTTS dataset, i.e., the speakers in the LibriTTS test set, the improvements of NMOS and SMOS increase to 7.02\% and 9.04\%, respectively, compared to YourTTS. Furthermore, adopting a more powerful TTS backbone accentuates these improvements when compared to Meta-StyleSpeech and StyleSpeech. Concurrently, we found that both USAT and YourTTS have a higher WER for the seen speakers of LibriTTS than StyleSpeech and Meta-StyleSpeech but a lower WER for all unseen speakers. We believe this is because the variational inference VITS framework is less prone to overfitting the training data, benefiting the model's generalization ability. Moreover, USAT also achieves the highest scores across all objective evaluation metrics. When considering the evaluation across datasets, i.e., speakers in the VCTK dataset, which can present a considerable generalization challenge to the model, USAT consistently outperforms all comparative models in every evaluated metric. 

\subsubsection{Evaluation in ESLTTS} \rec{We subsequently evaluate the instant adaptation for USAT and compare it with other zero-shot speaker-adaptive methods for speakers with heavy accents, i.e., speakers in ESLTTS. The results are presented in Table \ref{zsee}. Based on the results, we can observe that, through carefully designed disentangled representation learning between speaker information and linguistic information, USAT consistently outperforms all compared methods in heavy accent scenarios. \reaq{For instance, the SMOS of USAT is 6.97\% higher, than those of YourTTS.} Also, USAT surpasses all compared methods across all objective evaluation metrics.} However, by comparing with the evaluation results in VCTK, we notice that the naturalness and intelligibility of synthetic utterances remain similar, but the metrics related to speaker similarity, i.e., SMOS, SMCS, and SVR, show an \rec{evident} decline in all approaches. This decline underscores the challenges inherent in cloning the voices of heavily accented speakers for all approaches, illustrating the necessity of extending speaker-adaptive TTS evaluations beyond commonly utilized datasets, i.e., LibriTTS and VCTK, to thoroughly evaluate a model’s capabilities.

\begin{figure}[h]
\centering 
\subfloat[]{
\includegraphics[width=0.16\textwidth]{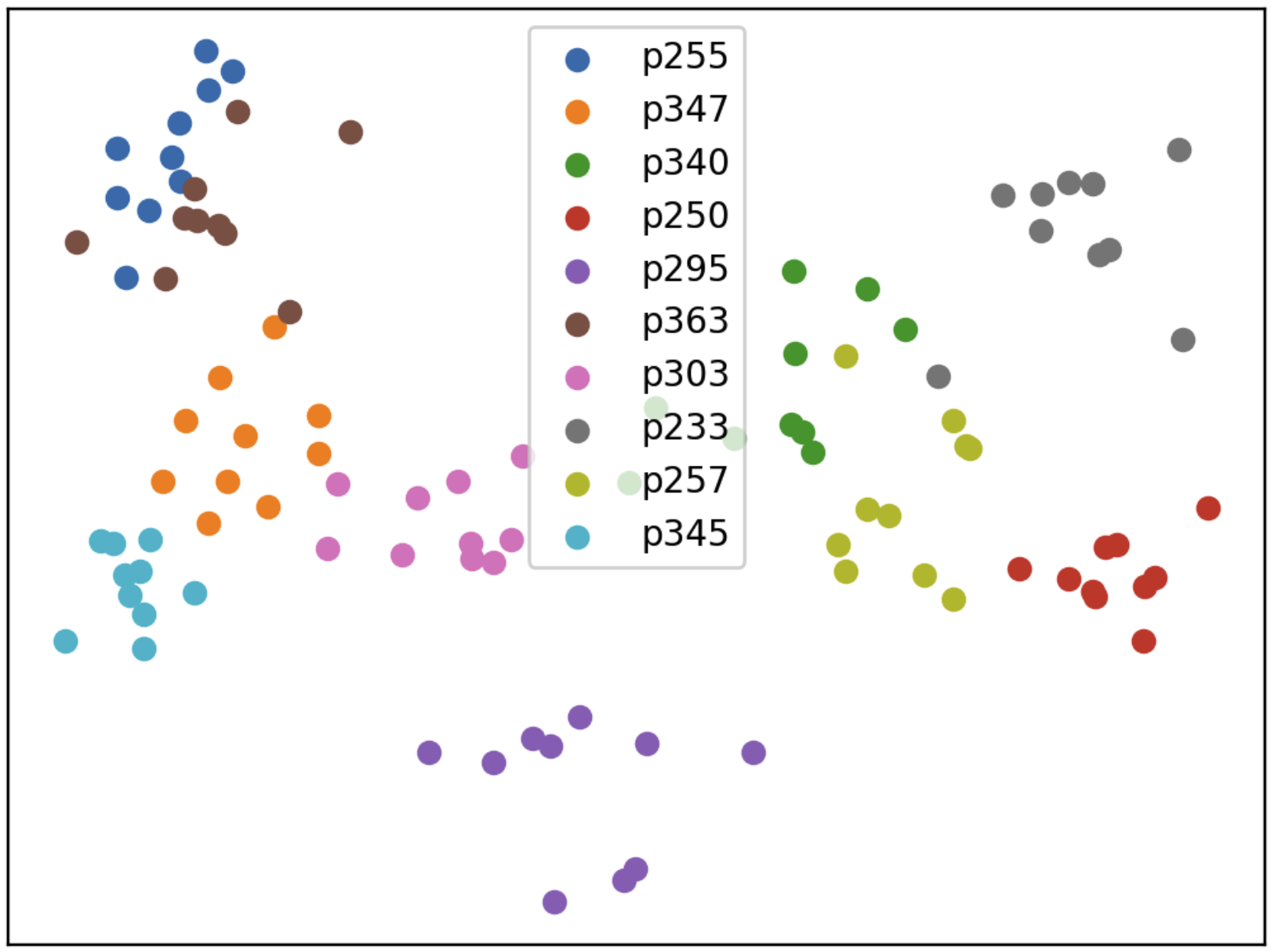}}
\hspace{3mm}
\subfloat[]{
\includegraphics[width=0.16\textwidth]{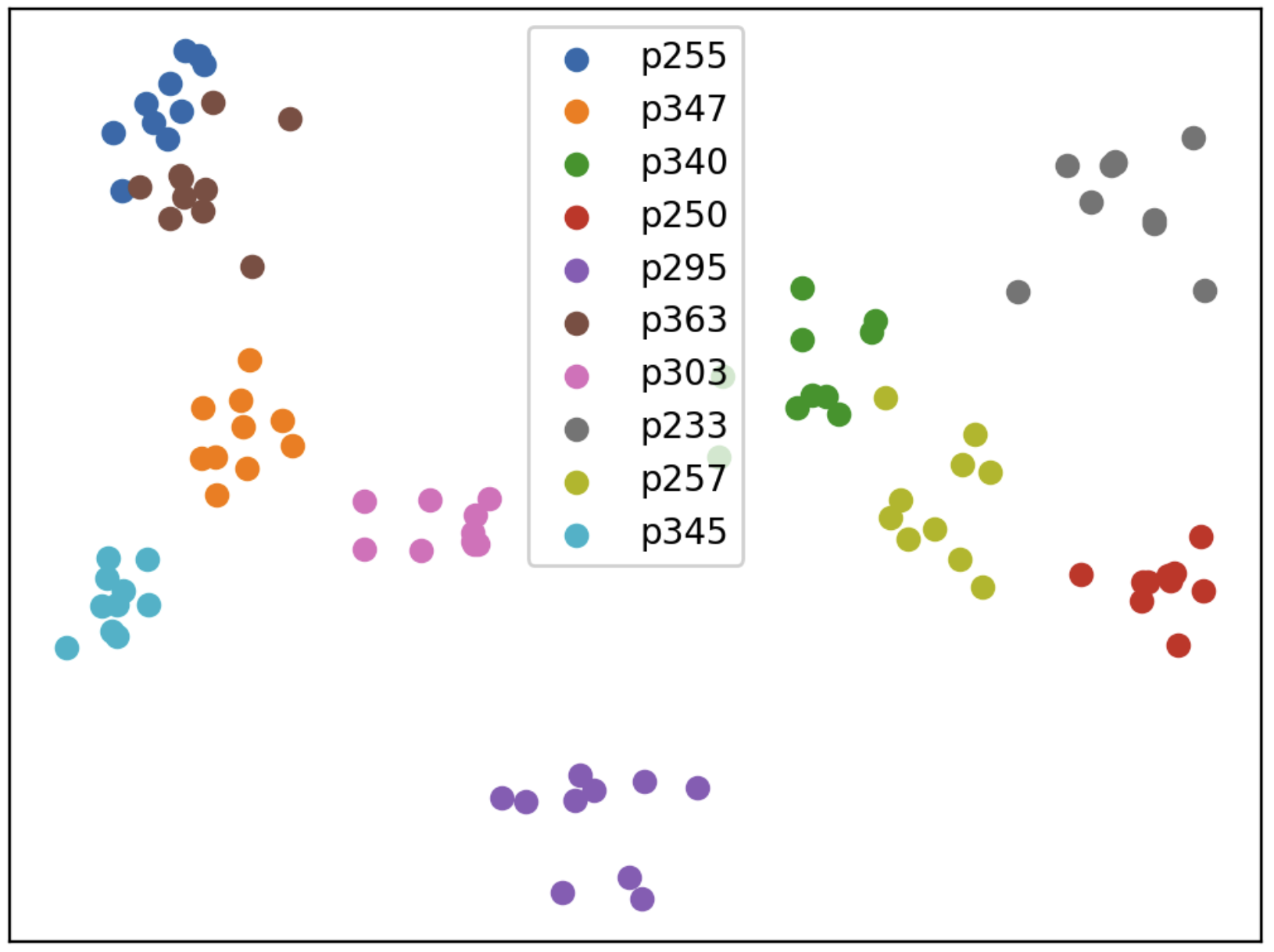}}
\caption{Visualization of speaker encoder embeddings after PCA dimensionality reduction (a) without phoneme leakage discriminator and (b) with phoneme leakage discriminator.}
\label{enc_wo}
\end{figure}

\begin{figure}[h]
\centering
\includegraphics[width=0.48\textwidth]{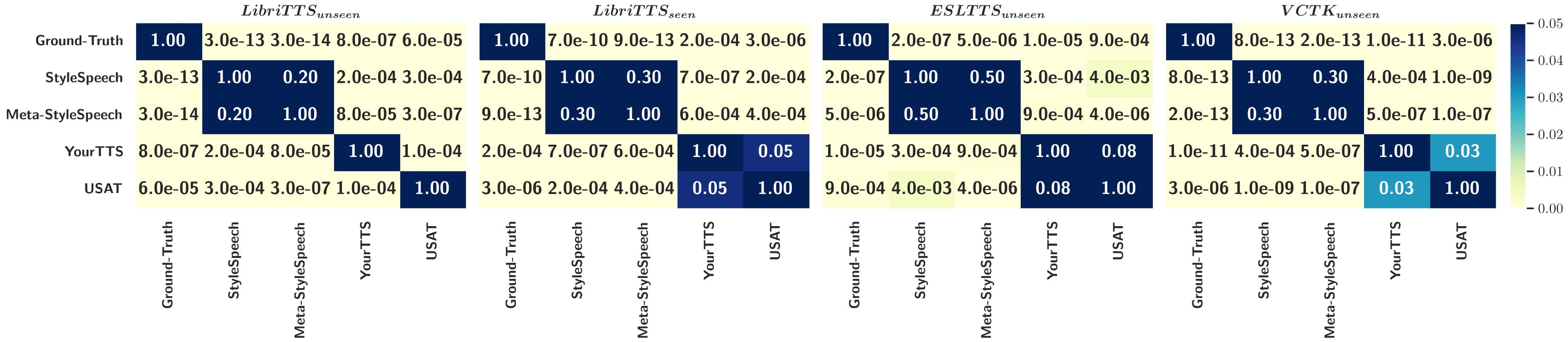}
\caption{\reaq{\(p\)-values of the Mann-Whitney U test for NMOS in instant speaker adaptation evaluation.}}
\label{npvalue}
\end{figure}

\begin{figure}[h]
\centering
\includegraphics[width=0.48\textwidth]{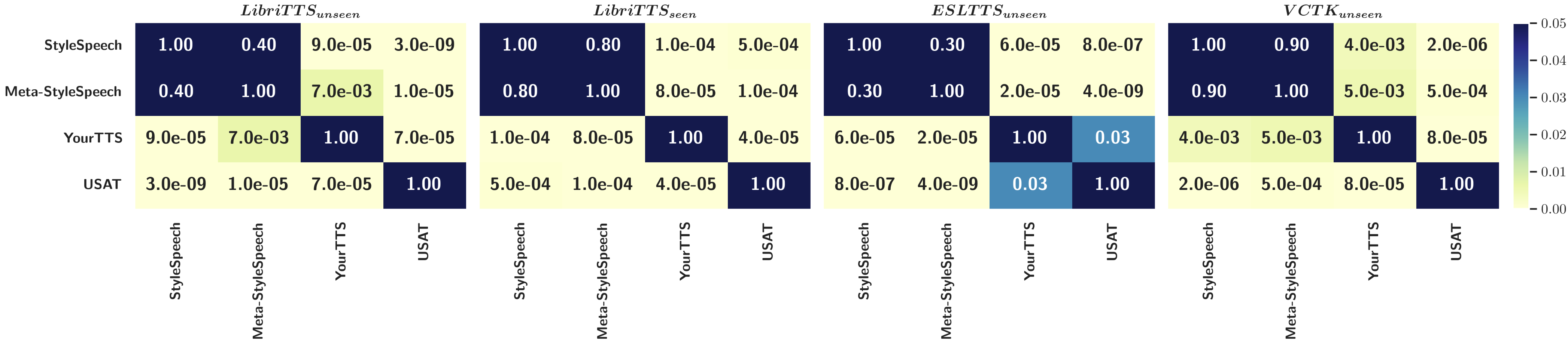}
\caption{\reaq{\(p\)-values of the Mann-Whitney U test for SMOS in instant speaker adaptation evaluation.}}
\label{spvalue}
\end{figure}

\reaq{
\subsubsection{Mann-Whitney U test of NMOS and SMOS evaluation results}
\label{utest}
To assess the statistical significance of NMOS and SMOS in subjective evaluations, we applied the Mann-Whitney U test, following \cite{DBLP:conf/interspeech/RosenbergR17, DBLP:conf/icassp/RosenbergFR18}. We utilized the two-sided Mann-Whitney U test implementation provided by SciPy \cite{2020SciPy-NMeth}. This non-parametric statistical method evaluates if differences in MOS scores between two models are statistically significant by comparing their rank sums, assessing the null hypothesis that a randomly chosen MOS result from one model is equally likely to be greater or smaller than the one from another model. Following \cite{wang2020asvspoof, DBLP:journals/access/SeshadriJRA19}, we used 0.05 as the significance level, i.e., \(p\)-value \(< 0.05\) suggests that the differences between the two models' MOS results are statistically significant. After pairwise applying the Mann-Whitney U test to the NMOS and SMOS results from various models and datasets, the \(p\)-values are shown in Figures \ref{npvalue} and \ref{spvalue}. For NMOS evaluations, we observed no statistically significant differences between StyleSpeech and Meta-StyleSpeech across all datasets. Likewise, there were no significant differences between YourTTS and USAT in the \(LibriTTS_{seen}\) and \(ESLTTS_{unseen}\). For the SMOS evaluation, the comparison between StyleSpeech and Meta-StyleSpeech on all datasets did not yield statistically significant results.
}

\begin{table}[h]
\caption{TTS ablation studies on \(\rm LibriTTS_{unseen}\).}
\begin{adjustbox}{width=8.5cm,center}
\label{ab}
\scriptsize
\begin{tabular}{c|ccccccc}
\toprule
Setting  & CNMOS      & CSMOS      & SMCS  & SVR(\%)    & UTMOS  &\(p\)-NMOS &\(p\)-SMOS \\ \midrule
USAT  & \(\rm0\)  & \(\rm0\) & \(\rm0.751\) & \(\rm80.1\) & \(\rm3.81\) & - & - \\ \midrule
w/o \(\mathcal{L}_{td}\) & \(\rm-0.01\) & \(\rm-0.13\)  & \(\rm0.744\) & \(\rm76.2\) & \(\rm3.80\) & \(0.914\) & \(0.039\) \\
w/o \(\mathcal{L}_{pd}\) & \(\rm-0.11\) & \(\rm-0.12\)  & \(\rm0.746\) & \(\rm71.0\)  & \(\rm3.76\) & \(0.057\) & \(0.048\) \\ 
w/o MAVAE & \(\rm-0.16\) & \(\rm-0.02\)  & \(\rm0.750\) & \(\rm79.1\)  & \(\rm3.72\)  & \(2e^{-4}\) & \(0.819\) \\ \bottomrule 
\end{tabular}
\end{adjustbox}
\end{table}

\reaq{
\subsubsection{Ablation Studies} 
We conducted ablation studies on USAT to assess the impact of each module, with results detailed in Table \ref{ab}. CNMOS and CSMOS represent the comparative NMOS and SMOS, respectively, highlighting differences with the default USAT's results. We also conducted the Wilcoxon signed rank test \cite{wilcoxon1992individual, DBLP:journals/corr/abs-2205-04421} on CNMOS and CSMOS to evaluate if the corresponding module statistically significantly enhances the model's score on NMOS and SMOS, with the corresponding \(p\)-value denoted as \(p\)-NMOS and \(p\)-SMOS. Removing the timbre residual discriminator (denoted as w/o \(\mathcal{L}_{td}\)) diminished SMCS, SVR, and SMOS scores. Then, excluding the phoneme leakage discriminator (denoted as w/o \(\mathcal{L}_{pd}\)) led to a decrease in SMOS, SVR, and UTMOS scores. We also observed a reduction in NMOS but it's not statistically significant. We believe this may be because ablation models do not exhibit substantially large MOS differences from the original model, and such differences are likely to be encompassed within the two models' confidence intervals, which is also observed in \cite{DBLP:conf/icml/CasanovaWSJGP22, DBLP:conf/interspeech/Wu00HZSQL22, DBLP:journals/corr/abs-2304-09116}, hence the observed non-highly significant p-values in the Mann-Whitney U test. We also conducted an ablation study on the VCTK dataset to probe the impact of the phoneme leakage discriminator on the speaker encoder. The visualization results in Fig. 9 demonstrate that the phoneme leakage discriminator improves the generalizability of the speaker encoder, reducing confusion and outliers in the extraction of embeddings for unseen speakers. Finally, removing the memory augmentation from the variational autoencoder (denoted as w/o MAVAE) resulted in a decrease in NMOS and UTMOS of the synthesized speech. This outcome is further mirrored in the training loss, i.e., the original model with memory augmentation can achieve a smaller sum of \(\mathcal{L}_{re} + \mathcal{L}_{kl}\) with larger \(\mathcal{L}_{kl}\) and smaller \(\mathcal{L}_{re}\), which suggests memory augmentation can help the VAE store more information in the latent space with the same dimension.
}
\begin{figure}[htp]
\vspace{-0.4cm}
\centering
\subfloat[]{
\label{conv_step}
\includegraphics[width=0.45\textwidth]{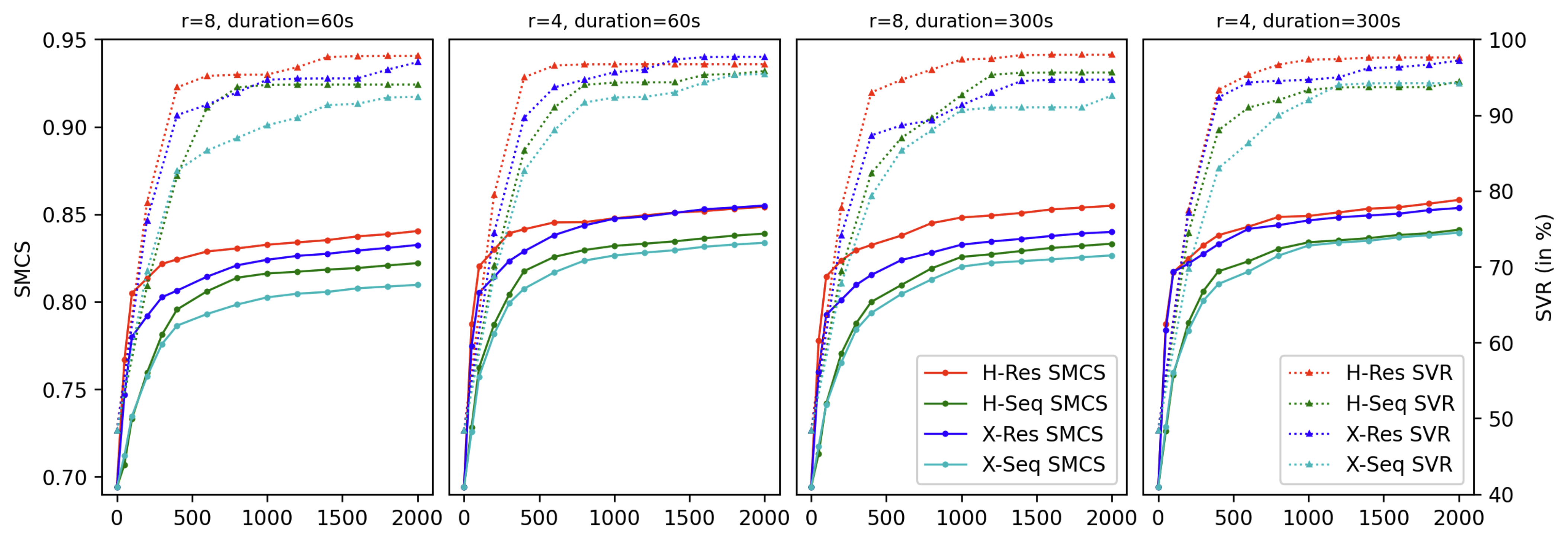}}
\\
\subfloat[]{
\label{lin_step}
\includegraphics[width=0.45\textwidth]{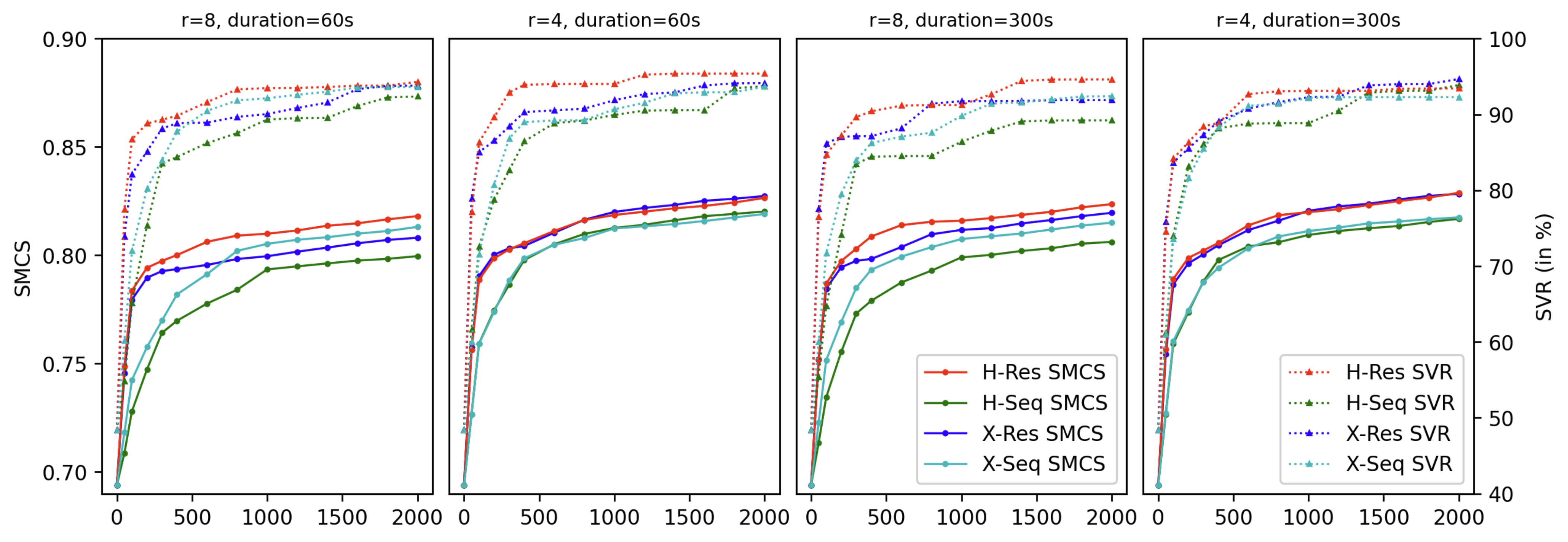}}
\caption{SMCS and SVR variation with the number of adaptation steps when employing (a) conv-flow adapters and (b) linear-flow adapters with different choices of \(r\), adaptation data duration, and adapter location during USAT's fine-grained adaptation. All figures share the same legends. The legend delineates the distinct adapter positions through line colors. Solid lines represent the evaluation metrics of SMCS, aligning with the left Y-axis. Conversely, dashed lines depict the evaluation metrics for SVR, corresponding to the right Y-axis.}
\vspace{-0.4cm}
\end{figure}

\begin{table}[]
\setlength\tabcolsep{3pt}
\caption{Evaluation results of retrained YourTTS.}
\begin{adjustbox}{width=8.5cm,center}
\label{retrain}
\begin{threeparttable}
\scriptsize
\begin{tabular}{c|cccc|cccc}
\toprule
\(\rm Dataset_{speaker}\)               & \multicolumn{4}{c|}{\(\rm LibriTTS_{unseen}\)}         &\multicolumn{4}{c}{\(\rm ESLTTS_{unseen}\)}\\ \midrule
Metric           & SMCS                      & WER(\%) & UTMOS  & SVR(\%)  & SMCS                    & WER(\%)    & UTMOS  & SVR(\%)            \\ \midrule
Ground-Truth         & 0.894     & 2.3    & 4.09 & 100   & 0.862                & 12.6  & 3.84 & 99.5   \\ \midrule 
YourTTS \cite{DBLP:conf/icml/CasanovaWSJGP22} & 0.702 & 6.0   & 3.69 & 67.8  & 0.674  & 14.1 & 3.68 & 32.7 \\
YourTTS (Retrained) &  0.723  &  \boldmath\(5.8\)   & 3.73 & 72.9  & 0.682  & \textbf{14.0} & 3.71 & 38.2 \\
USAT          & \textbf{0.751} & 5.9   & \boldmath\(3.81\) & \boldmath\(80.1\)  & \textbf{0.694}   & \textbf{14.0} & \textbf{3.82} & \textbf{48.4}\\ \bottomrule
\end{tabular}
\end{threeparttable}
\end{adjustbox}
\end{table}

\reaq{
\subsubsection{Retraining YourTTS in LibriTTS} Since YourTTS was originally trained across various datasets in multiple languages with a language-based batch balancer, there is a risk of overfitting on the vocal characteristics of languages represented by a limited number of speakers. For example, even though the Portuguese training dataset only features a single speaker, according to the language-based batch balancer, it accounts for one-third of the total training data, which could lead to the model being overfitting to this particular speaker's voice. To facilitate a fair comparison between USAT and YourTTS, we retrained YourTTS on the ``train-clean-100'' and ``train-clean-360'' subsets of the LibriTTS dataset following the original training configuration. Subsequently, we evaluated the retrained YourTTS and USAT models on \(LibriTTS_{unseen}\) and \(ESLTTS_{unseen}\) datasets, with results shown in Table \ref{retrain}. According to the results, the retrained YourTTS model exhibited improved performance on both \(LibriTTS_{unseen}\) and \(ESLTTS_{unseen}\) datasets compared to its original version, suggesting potential overfitting in the pre-trained YourTTS checkpoint. Notably, retrained YourTTS surpassed USAT in the WER metric on the \(LibriTTS_{unseen}\) dataset, which may be attributed to the additional Transformer \cite{DBLP:conf/nips/VaswaniSPUJGKP17} blocks in YourTTS's phoneme encoder, which is initially designed for the multilingual TTS. Besides the WER, USAT outperforms YourTTS across most evaluation metrics. We believe that in addition to the improvements we implemented in the VITS backbone, this is another significant difference between USAT and YourTTS contributing to its superiority: unlike YourTTS, which utilizes speaker embeddings pre-trained on discriminative tasks, USAT employs learnable speaker embeddings and is trained as a generative task. This approach favors the zero-shot speaker-adaptive TTS task, as corroborated by \cite{DBLP:conf/icassp/ChienLHHL21}.
}

\subsection{Fine-grained Adaptation Evaluation}
\subsubsection{The Variables in Fine-grained Adaptation}
\label{adavar}
To investigate the influence of variables like adapter placement on fine-grained adaptation, as described in Section \ref{fsada}, we conducted evaluations focusing on various adapter configurations with an extra set of 10 speakers from the ESLTTS dataset (no overlap with the evaluation set of 30 speakers). Due to the absence of significant differences in the naturalness of the synthetic speech, our primary emphasis remains on metrics corresponding to speaker similarity, i.e., SMCS and SVR. The evaluation results for the conv-flow adapter are depicted in Fig.\ref{conv_step}, while the results of the linear-flow adapter can be found in Fig.\ref{lin_step}. Several interesting observations are made from these results: Concerning the adapter's insertion locations and methods, both the conv-flow and linear-flow adapters exhibit faster and superior adaptation with a residual insertion (denoted as \(*-Res\) ) compared to a sequential one (represented as \(*-Seq\)). Furthermore, in most scenarios, inserting the adapter within the transformation function (denoted as \(H-*\)) tends to produce enhanced adaptation outcomes in contrast to its placement adjacent to the transformation function (denoted as \(X-*\)). Concerning the hyperparameter \(r\), a heightened sensitivity is observed when the adapter is adjacent to the transformation function. More explicitly, an \(r\) value of 4 offers similar adaptation results irrespective of the adapter's \rec{insertion} methods. However, doubling \(r\), or halving the adapter's bottleneck dimension equivalently, leads to a notable performance drop for adapters adjacent to the transformation function. This degradation is not observed when the adapter is inside the transformation function. Concerning the data duration for adaptation, there is a negligible disparity in performance between adapters trained on the 60-second subset and the original 300-second set from the ESLTTS dataset, and this suggests that a duration of 60 seconds of reference data is sufficient for flow adapters to learn all the features of the target speaker's voice. Concerning the number of adaptation steps, while the SMCS metric already performs well, approximately at 500 steps, the SVR index typically experiences volatility until around 1000 steps, stabilizing subsequently at about 1500 steps. This result indicates that flow adapters usually need 1500 adaptation steps to capture the target speaker’s voice characteristics comprehensively.

\begin{table}[t]

\setlength\tabcolsep{3pt}
\caption{Evaluation Results of Few-Shot Speaker Adaptation in ESLTTS.}
\begin{adjustbox}{width=8.8cm,center}
\label{fse}
\begin{threeparttable}
\scriptsize
\begin{tabular}{c|ccccccc}
\toprule
\(\rm Dataset_{speaker}\)     & \multicolumn{7}{c}{\(\rm ESLTTS_{unseen}\)}           \\ \midrule
Metric           & NMOS             & SMOS            & SMCS  & WER(\%) & UTMOS  & SVR(\%)  & \#Para(M)            \\ \midrule
Ground-Truth     & \(4.26\pm0.09\) & -               & 0.862   & 12.6   & 3.52 & 99.5  & - \\ 
USAT(Instant)    & \(3.86\pm0.08\) & \(3.22\pm0.09\) & 0.694   & 14.0   & 3.48 & 48.4  & 0.00 \\\midrule 
VITS(Full) \cite{DBLP:conf/icml/KimKS21}     & \boldmath\(3.80\pm0.07\) & \(3.56\pm0.07\) & 0.795   & 14.6   & 3.40 & 84.8  & 36.4 \\
UnitSpeech \cite{DBLP:journals/corr/abs-2306-16083}    & \boldmath\(3.84\pm0.09\) & \boldmath\(3.75\pm0.06\)  & \textbf{0.834}   & \textbf{10.3}   & 3.49 & \textbf{99.0}  & 119.1 \\
USAT(Fine-grained)    & \boldmath\(3.84\pm0.07\) & \boldmath\(3.74\pm0.05\) & 0.833   & 11.0   & \textbf{3.52} & 98.8  & \textbf{0.64} \\ \bottomrule
\end{tabular}
\end{threeparttable}
\end{adjustbox}
\end{table}

\begin{table}[t]
\setlength\tabcolsep{3pt}
\caption{Few-Shot Speaker Adaptation Ablation Study.}
\begin{adjustbox}{width=8.8cm,center}
\label{aba}
\begin{threeparttable}
\scriptsize
\begin{tabular}{c|ccccccc}
\toprule
\(\rm Dataset_{speaker}\)     & \multicolumn{7}{c}{\(\rm ESLTTS_{unseen}\)}           \\ \midrule
Metric           & NMOS             & SMOS            & SMCS  & WER(\%) & UTMOS  & SVR(\%)  & \#Para(M)            \\ \midrule
USAT(Modules)    & \boldmath\(3.83\pm0.09\) & \boldmath\(3.75\pm0.06\)  & \textbf{0.835}   & 13.4   & 3.44 & \textbf{99.0}  & 16.3 \\
USAT(Fine-grained)    & \boldmath\(3.84\pm0.07\) & \boldmath\(3.74\pm0.05\) & 0.833   & \textbf{11.0}   & \textbf{3.52} & 98.8  & \textbf{0.64} \\ \bottomrule
\end{tabular}
\end{threeparttable}
\end{adjustbox}
\end{table}
\begin{table}[t]
\setlength\tabcolsep{3pt}
\caption{Few-Shot Speaker Adaptation Evaluation Results of YourTTS and USAT.}
\begin{adjustbox}{width=8cm,center}
\label{ffyour}
\begin{threeparttable}
\scriptsize
\begin{tabular}{c|ccccc}
\toprule
\(\rm Dataset_{speaker}\)     & \multicolumn{5}{c}{\(\rm ESLTTS_{unseen}\)}           \\ \midrule
Metric             & SMCS  & WER(\%) & UTMOS  & SVR(\%)  & \#Para(M)            \\ \midrule
YourTTS(w weighted-sampling)   & 0.809   & 11.5   & 3.46 & 88.1  & 40.1 \\
YourTTS(w/o weighted-sampling)   & 0.821   & 13.2   & 3.44  & 93.2  & 40.1 \\
USAT(Fine-grained)    & \textbf{0.833}   & \textbf{11.0}  & \textbf{3.52} & \textbf{98.8}  & \textbf{0.64} \\ \bottomrule
\end{tabular}
\end{threeparttable}
\end{adjustbox}
\end{table}

\subsubsection{Evaluation in ESLTTS} 
We compared the performance of USAT's fine-grained adaptation with several other few-shot-based approaches on the ESLTTS dataset. The outcomes are illustrated in Table \ref{fse}, where the column labeled “\#Para(M)” represents the number of parameters in millions (M) used for fine-tuning. For the fine-grained adaptation of USAT, we set \(r\) to 8 and employed the convolution-layer-based \(H-Res\) adapter. The adaptations are uniformly executed on the 60-second adaptation subsets, as described in Section \ref{adavar}. For both UnitSpeech and USAT(Fine-grained), the adaptation encompassed 1500 steps. In contrast, for other adaptation approaches, we adapted them for more than 1500 steps to achieve the best quality. Empirical observations reveal that the fine-grained adaptation of USAT can offer comparable or even superior adaptation outcomes relative to other approaches by fine-tuning only 0.5\% to 1.6\% amount of the parameters compared to other methods. \reaq{We also performed the Mann-Whitney U test on the evaluation results of fine-grained adaptation for NMOS and SMOS. The test revealed that there were no statistically significant differences in the NMOS across all compared methods of speech synthesis, i.e., all \(p\)-values between them \(> 0.05\). Similarly, the \(p\)-value between UnitSpeech's and USAT's (Fine-grained) SMOS scores is 0.6, indicating there is no statistically significant difference.}

\reaq{
\subsubsection{Comparison with YourTTS few-shot speaker adaptation} We also compared USAT with YourTTS in the fine-grained adaptation scenario. Notably, YourTTS incorporates weighted random sampling during the adaptation stage, ensuring that one-quarter of the samples in each training batch are from the target speaker, with the remainder drawn from various other speakers. This differs from our approach and the other methods we evaluated. Therefore, we compared USAT with two variations of YourTTS: one employing the original weighted random sampling and another without it, where all adaptation samples are exclusive from the target speaker. As shown in Table \ref{ffyour}, weighted random sampling effectively prevents overfitting on limited speech data, reflected in improved WER and UTMOS scores. In contrast, adapting the model directly with data from the target speaker, without weighted sampling, results in better speaker similarity, indicated by higher SMCS and SVR scores. Moreover, USAT outperforms both versions of YourTTS while using fewer parameters, suggesting that our adaptor approach may offer a better solution for fine-grained speaker adaptation.
}

\subsubsection{Adapting Adapters vs. Adapting Whole Relevant Modules} \rec{To evaluate whether only adapting the adapters can achieve equivalent synthetic speech quality compared to fine-tuning the whole speaker-relevant modules without adapters, i.e., the timbre flow, phoneme encoder and duration predictor, we conducted an ablation experiment. The results are shown in Table\ref{aba}. Compared with fine-tuning all speaker-related modules, adjusting only the adapter can achieve similar naturalness and speaker similarity by fine-tuning less than 4\% of the parameters.}

\begin{figure}[htp]
\centering
\subfloat[]{
\includegraphics[width=0.23\textwidth]{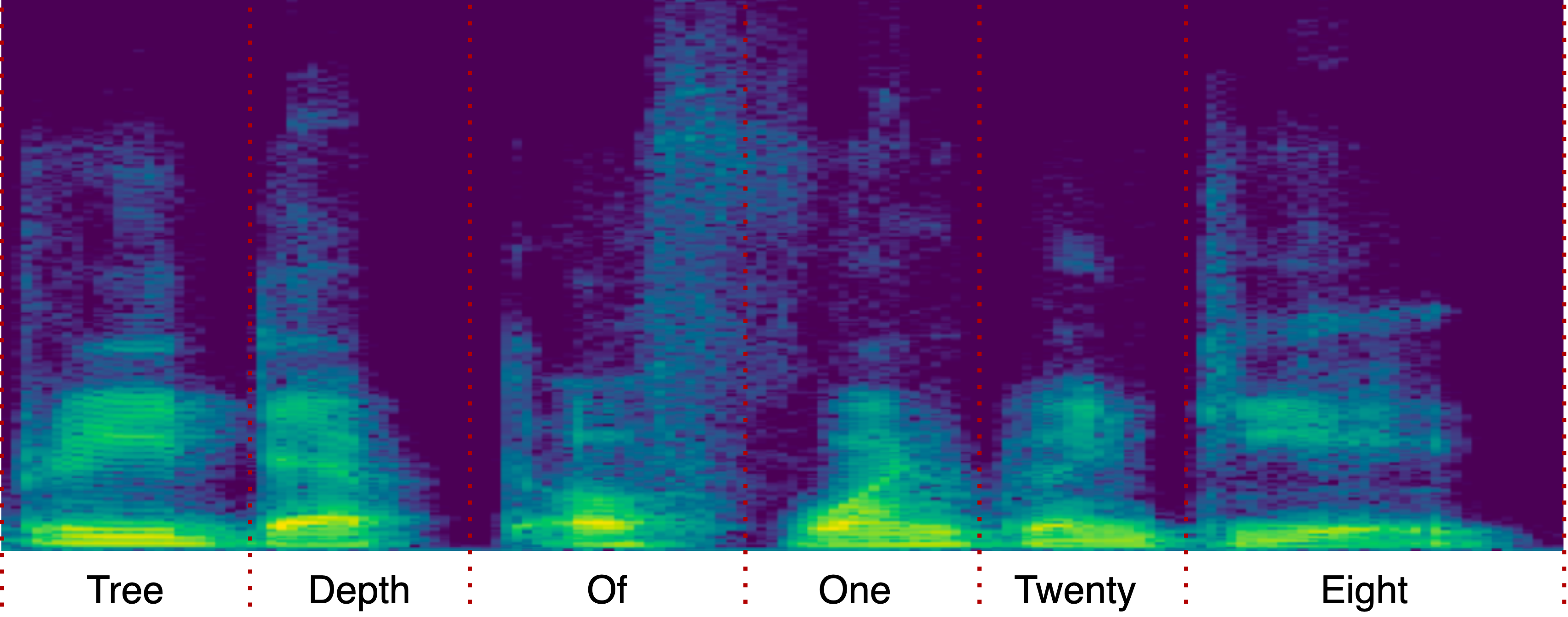}}
\subfloat[]{
\includegraphics[width=0.245\textwidth]{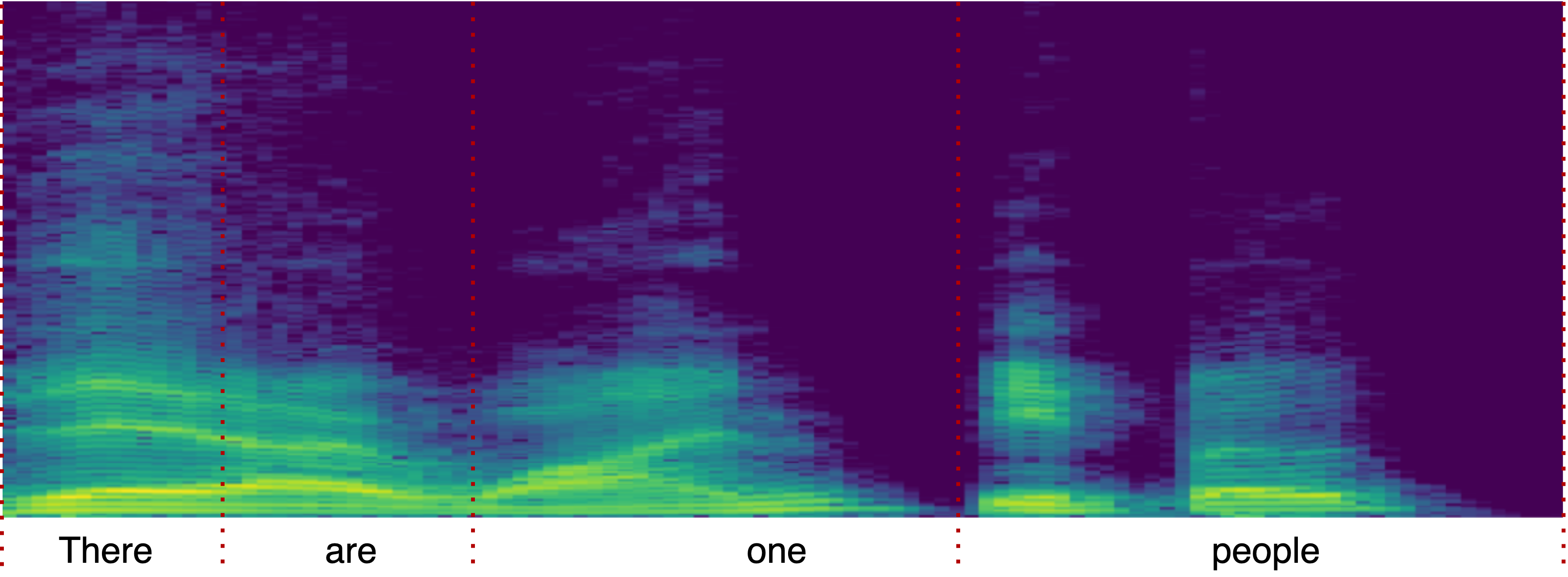}}
\\
\subfloat[]{
\includegraphics[width=0.24\textwidth]{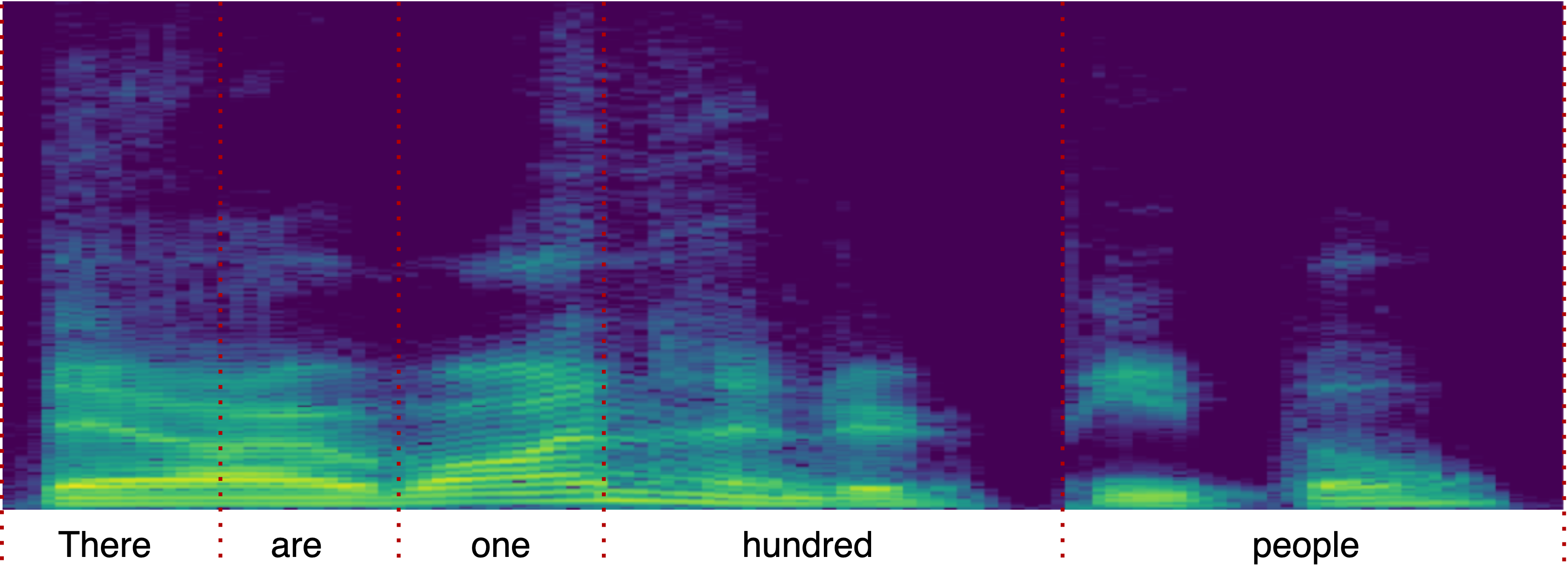}}
\caption{\label{overfit} (a) Spectrogram of the reference utterance, where ``one hundred and twenty eight'' is mispronounced as ``one twenty eight''. (b) Spectrogram of \rec{the} synthetic utterance (adaptation whole model) with the text ``There are one hundred people'', we can find that the model omits the word ``hundred'' due to catastrophic forgetting. (c) \rec{The} synthetic utterance spectrogram (adaptation adapter) with the text ``There are one hundred people'', we can find that the model is resistant to the problem of catastrophic forgetting and pronounces the word ``hundred'' correctly. Corresponding utterances can be found in our online demo.}
\end{figure}

\subsubsection{Do Adapters Offer Resistance to Overfitting and Catastrophic Forgetting?} To evaluate the resilience of our designed adapters to overfitting and catastrophic forgetting during fine-grained adaptation, we analyzed the speech synthesized during our experiments. A representative example is illustrated in Fig. \ref{overfit}. \rec{Suppose an inconsistency arises between the waveform speech and the text of a reference utterance in the adaptation dataset. For instance, while the reference text reads ``one hundred twenty-eight'', the actual waveform speech says ``one twenty-eight'', with the word ``hundred'' omitted, as depicted in Fig. \ref{overfit}a. We observed a marked difference in the synthesized speech when comparing the results of adapting the USAT’s modules to those of solely adapting the adapters. Specifically, upon adapting the USAT’s modules, the adapted model forgot how to pronounce the word “hundred” when synthesizing speech from the text “There are one hundred people”, as evident in Fig. \ref{overfit}b. Conversely, by only adapting the adapters, this forgetting was avoided. The word “hundred” was correctly pronounced with the same acoustic feature as the target speaker, as shown in Fig. \ref{overfit}c.} The corresponding utterances can be accessed in our online demo.

\section{CONCLUSION}
\label{sec:con}
We introduced USAT, a novel universal speaker-adaptive TTS framework, offering two distinct adaptation strategies to tackle challenges in synthesizing speech for unseen speakers in the wild. Key innovations include specialized discriminators for improved generalization and unique adapters to address storage and overfitting concerns. We also proposed the ESLTTS dataset, featuring diverse non-native English accents, facilitating more holistic speaker-adaptive TTS evaluations and research. Comparative studies demonstrate USAT’s superior performance over existing methodologies across multiple metrics. In future work, we plan to further improve the generalizability of zero-shot speaker-adaptive TTS and explore more adaptation approaches for few-shot speaker-adaptive TTS, such as the adaptive factorized weight mechanism \cite{DBLP:conf/icassp/NguyenPW23}.

\vfill

\end{document}